\documentclass[conference]{IEEEtran}
\pdfoutput=1
\IEEEoverridecommandlockouts
\usepackage{booktabs} 
\usepackage{subfigure}

\usepackage{amsmath}
\usepackage{amssymb}
\usepackage{amsthm}
\usepackage{math tools}
\usepackage[normalem]{ulem}
\usepackage{dutchcal}
\usepackage{tabularx}
\usepackage{enumerate}
\usepackage{authblk}

\newcommand{\stitle}[1]{\vspace*{0.5em}\noindent{\bf #1\/} }

\newcommand{\ya}{\ensuremath{t^j_i}}
\newcommand{\yb}{\ensuremath{l^j_i}}
\newcommand{\w}{\mathsf{w}}

\newcommand{\grad}{\ensuremath{\Delta}}

\newcommand{\loss}{\ensuremath{\mathbcal{l}}}
\newcommand{\T}{\ensuremath{T}}
\newcommand{\name}{TensorFlowOnline}

\newcommand{\best}{{\sf Best}~}
\newcommand{\worst}{{\sf Worst}~}

\begin{document}

\title{Towards Self-Tuning Parameter Servers}

\author[*]{Chris Liu}
\author[*]{Pengfei Zhang}
\author[**]{Bo Tang}
\author[***]{Hang Shen}
\author[*]{Lei Zhu}
\author[*]{Ziliang Lai}
\author[*]{Eric Lo}
\affil[*]{The Chinese University of Hong Kong}
\affil[**]{Southern University of Science and Technology}
\affil[***]{SiChuan University}

\maketitle
\begin{abstract}
Recent years, many applications
have been driven advances by the use of Machine Learning (ML).
Nowadays,
it is common to see industrial-strength machine learning jobs that involve millions of model parameters, terabytes of training data, and weeks of training.
Good efficiency, i.e., fast completion time of running a specific ML job, therefore, is a key feature of a successful ML system.
While the completion time of a long-running ML job is determined by the time required to reach model convergence,
practically that is also largely influenced by the values of various system settings.
In this paper, we contribute techniques towards building \emph{self-tuning parameter servers}.
Parameter Server (PS) is a popular system architecture for large-scale machine learning systems;
and by self-tuning we mean
while a long-running ML job is iteratively training the expert-suggested model,
the system is also iteratively learning which system setting is more efficient for that job and
applies it online.
While our techniques are general enough to various PS-style ML systems, 
we have prototyped our techniques
on top of {\sf TensorFlow}.
Experiments show that our techniques can reduce the completion times of 
a variety of long-running {\sf  TensorFlow} jobs from 1.4$\times$ to 18$\times$.
\end{abstract}

%
%
%
%


\section{Introduction}
\label{intro}

Recently, the \emph{Parameter Server} (PS) architecture
\cite{DeanCMCDLMRSTYN12nips, XingHDKWLZXKY15kdd, LiAPSAJLSS14osdi, MxNet,adam,tensorflow}
has emerged as a popular system architecture to support large-scale distributed learning
on a cluster of machines.  The PS architecture advocates the separation of working units as ``servers'' and ``workers'', where the servers collectively maintain the model state and the workers duly ``pull'' the latest version of the model  from the servers,
scan their own part of training data to compute the model refinements, and ``push'' the model updates back to the servers for aggregation.
The PS architecture has the advantage of improving the network utilization so that it can scale-out to bigger model and more machines.

%
%
Generally,
the completion time of a long-running machine learning job is determined by the time required to reach model convergence. Practically, however, the completion time is largely influenced by the values of the various system knobs, such as  the server-worker ratio
(i.e., how many hardware threads are dedicated to the servers and workers),
the device placement (i.e., which operation shall be shipped to the GPU for processing and which shall stay in the CPU),
and the parallelism degree (e.g., the model replication factor, the model partitioning scheme).
Today, unfortunately, the burden falls on the users who submit the ML jobs to specify the knob values.

\begin{figure}\centering
	\includegraphics[width=8cm]{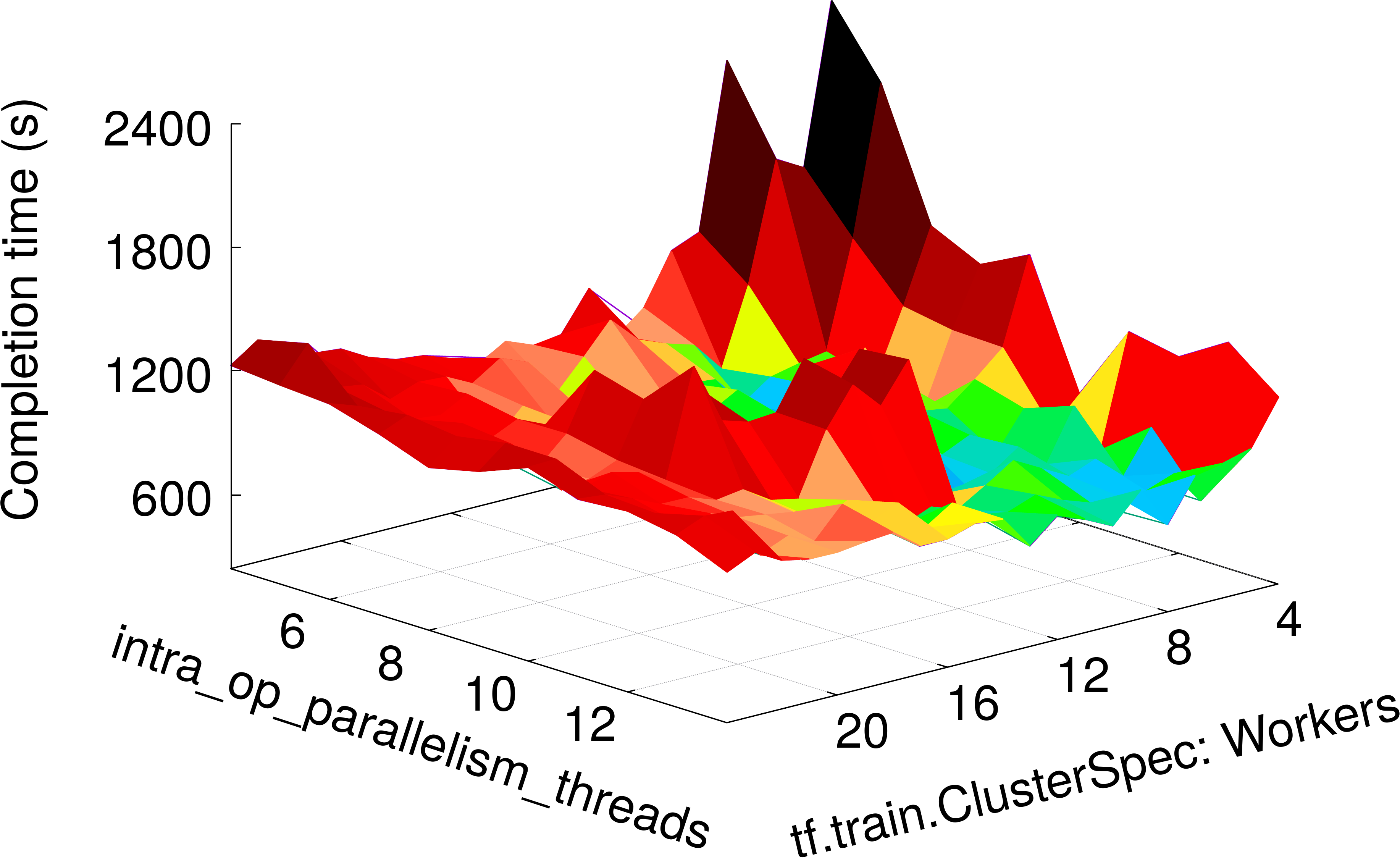}
	\caption{A 2D response surface of a  {\sf TensorFlow} job}
	\label{fig:3d}
\end{figure}

\begin{figure}\centering
	\includegraphics[width=8cm]{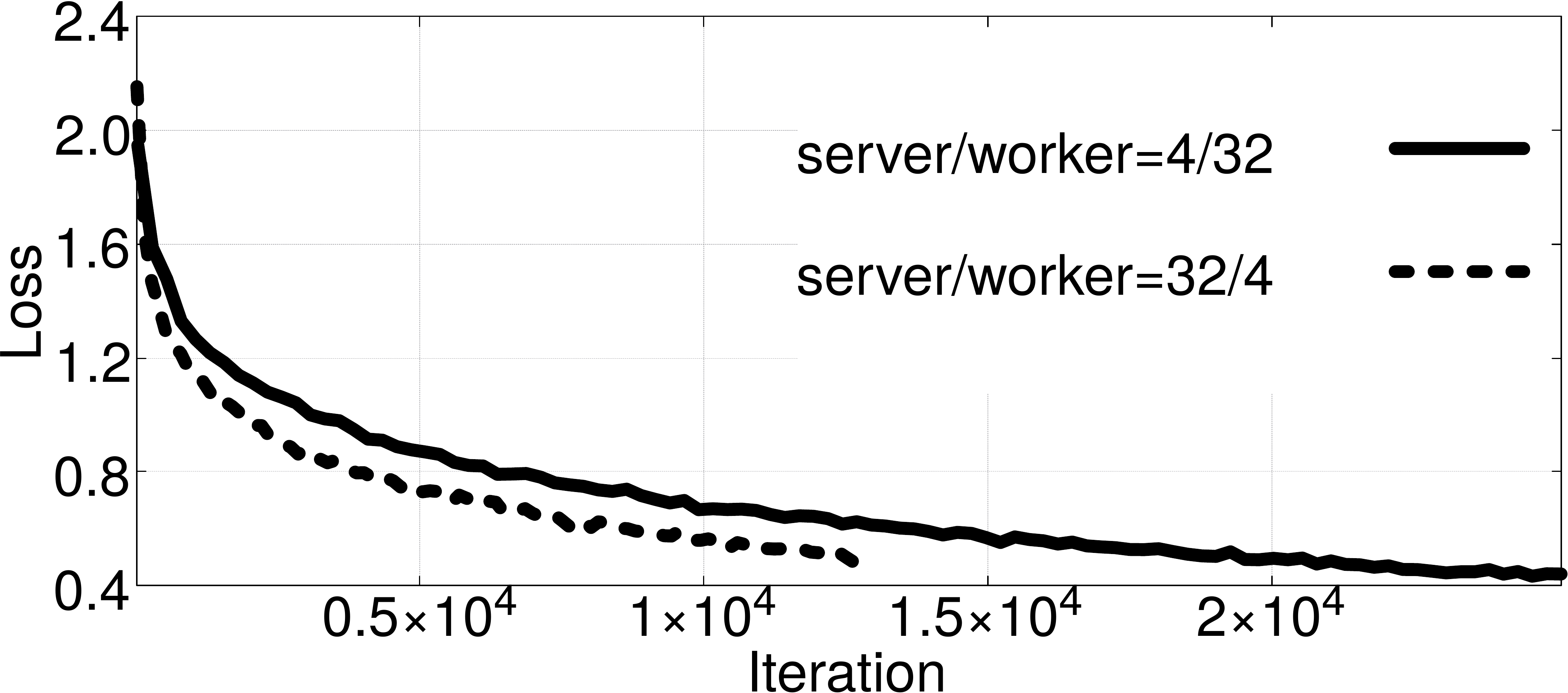}
	\caption{Different system settings would influence statistical efficiency (convergence threshold = loss 0.4)}
	\label{fig:2d}
\end{figure}

Determining the right set of knob values that achieve optimal completion time has way surpassed human abilities. Part of what makes that so enigmatic is that the
response surfaces of ML jobs are highly complex.
Figure \ref{fig:3d} shows a response surface of running a 
PS-style {\sf TensorFlow} job on our cluster (experiment details are presented later).
The two system knobs involved are: \emph{tf.train.ClusterSpec::worker} and \emph{intra\_op\_parallelism\_threads},
which respectively vary  the server-worker ratio
and the thread affinity of operations (i.e., the mapping between {\sf TensorFlow} operations and hardware threads).
The figure shows that the response surface is complex and non-monotonic,
and the optimal lies at where human can't easily find.
What adds to the challenge is that,
the completion time of a ML job, unlike traditional data processing,
is a complex interplay between \emph{statistical efficiency} (how many iterations are needed until convergence) and \emph{hardware efficiency} (how efficiently those iterations are carried out) \cite{ZhangR14vldb}.
Consider the server-worker ratio as an example.  On the one hand, more workers would increase the hardware efficiency by having a higher degree of data parallelism.  On the other hand, that might hurt the statistical efficiency when servers accept asynchronous updates from workers.
That is because when more  workers concurrently update the global model, the model would be more inconsistent and require more iterations to converge.
Figure \ref{fig:2d} shows such a case.
It shows that varying just one system setting (server-worker ratio) would already yield a 2.5$\times$ difference in statistical efficiency 

Configuring distributed ML systems to reduce the long-running execution time of ML jobs currently requires system expertise -- something many ML users may lack.
Even for system experts, the dependencies between the knobs (e.g, changing one knob may nullify the benefits of another)
make the whole task nontrivial if that is not downright impossible.
Furthermore, this manual tuning task must be repeated whenever the expert-suggested model or hardware resources changes.

In this paper we present a suite of techniques for building \emph{self-tuning parameter servers}.
By self-tuning, we mean
when a long-running ML job is iteratively training an expert-suggested model,
the system itself is also iteratively learning which setting is more efficient for that job and
dynamically applies it online.
Our goal is to free ML users from the system details as much as they could and
let the system progressively discover and apply better and better system settings for a job as it proceeds.
The principled contributions of this paper are as follows:

\begin{enumerate}
\item {\sf Online Optimization Framework}  (Section \ref{tuning}).
We present a framework that is applicable to all PS-style ML systems to support self-tuning.
In online tuning,
we always hope to discover the optimal system setting as soon as possible
(so as to apply that in the next iteration immediately)
while minimizing the number of iterations to discover it.
Therefore, there is an intrinsic balance between
trying a potentially better system setting
or applying a known good setting before the start of each iteration.
The crux of the framework is the use of \emph{Bayesian Optimization} (BO) \cite{BOTutorial, BONIPS2012, BAGO}
to learn and recommend a system setting.
Bayesian Optimization has been extensively used in various offline auto-tuning projects (e.g., \cite{easeml}, \cite{cherrypick}).
However, there are subtleties 
when coming to online ML system tuning.
For example, 
%
in ML learning, the influence of the same system setting would depend on whether the training just starts or the model is converging.
%
Therefore, we have formulated a BO that is aware of those intrinsic ML factors.

\item {\sf Online Progress Estimation} (Section \ref{progress}).
A key input to our Bayesian Optimization 
is the \emph{estimated remaining completion time} of an \emph{in-flight} ML job.
The challenge there is to estimate the \emph{statistical progress} --- ``how many \emph{more} iterations to go based on  the current model and current system setting?''.
While there is a wealth of works that study the convergence rates of different learning algorithms on different ML problems \cite{RechtRWN11nips, Tamingthewild, ConvergenceAnalysis}, they are all theoretical bounds under an \emph{offline} setting,
i.e., bounding the maximum number of iterations required given an untrained model. 
%
%
%
%
%
%
%
%
%
%
%
%
Focusing on gradient descent based learning algorithms (e.g., SGD, Hogwild! \cite{RechtRWN11nips}),
our contributions here are a formal extension of those bounds for an online setting
and a methodology to transform those bounds to be legitimate statistical progress estimation functions.

\item {\sf Online Reconfiguration} (Section \ref{reconfig}).
None of the above would be meaningful 
unless there is a way to  online reconfigure a PS-style ML system to a new system setting while active jobs are still running.
While most ML systems have built-in checkpointing facilities  for recovery purpose,
implementing online reconfiguration by checkpointing the state of a job 
and restoring the state under the new system setting would incur excessive overhead and cause system quiescence.  
To this end, we introduce a novel technique called \emph{On Demand Model Relocation} (ODMR) 
so that non-quiescent and efficient online reconfigurations can be carried out.

\item {\sf Experimentation on {TensorFlow}} (Section \ref{exp}).
One point worths noticing is that all of the  our  contributions (online optimization framework, online progress estimation, and online reconfiguration) above
are system agnostic.
That is, any existing PS-style systems can implement our techniques and enjoy faster end-to-end runtime through self-tuning.
As an experimental prototype,
we have implemented our techniques on top of {\sf TensorFlow} \cite{tensorflow} and we name it  {\sf \name}.
Experiments show that {\sf \name} can
reduce the long-running completion times of different {\sf TensorFlow} jobs by 1.4$\times$--18$\times$.
{\sf \name} is open-source. So its statistical progress estimation component can be easily adapted to any new convergence results  from the machine learning community.

%
%

\end{enumerate}

%
%

In this paper, we focus on online {\bf system parameter} tuning.
By system parameters we refer to those would influence only the efficiency but {\bf not} the quality of the models.  Therefore, the server-worker ratio in {\sf Tensorflow} is a system parameter
while the learning batch size is not.  Instead, the learning batch size is a \emph{hyperparameter}
because it influences the quality of the final model and is not learnable from the data.
With this difference is clear, we can see that 
our work is orthogonal to projects that focus on hyperparameter tuning (e.g., Spark TuPAQ \cite{SparksTFJK15}, Auto-WEKA \cite{KotthoffTHHL17}, Google Vizier \cite{GolovinSMKKS17}, Spearmint \cite{Becker-KornstaedtSZ00}, GpyOpt\cite{gpyopt2016}, and Auto-sklearn \cite{autosklearn}, Ease.ml \cite{easeml}). 
In fact, hyperparameter tuning is generally a trial-and-error process,
where each trial is a different job using a different set of hyperparameters (e.g., learning rate, batch size) under the same system setting (e.g., worker-server ratio). 
Our work thereby can complement those systems to expedite the execution of each trial
and shorten the overall hyperparameter tuning cycle.

Next, we present the preliminary and background for this paper (Section \ref{background}),
followed by our main contributions (Sections \ref{tuning}--\ref{exp}).
We give a review of related work afterwards (Section \ref{related})
and conclude this paper with some interesting future directions  (Section \ref{conclusion}).
The appendix includes some detailed implementation of our prototype \name, additional experimental results, and a table that summarizes the major notations used in this paper.

\section{Background and Preliminary}
\label{background}

ML jobs come in many forms, such as
logistic regression,
support vector machine, 
and neural networks.
Nonetheless, almost all seek a model 
$w$ of parameters that
minimize an \emph{empirical risk function} $R$:

$$R(w) = \frac{1}{n}\sum_{i=1}^{n}l(w, d_i)$$
where $n$ is the number of data examples in the whole training dataset $\mathcal{D}$ 
and $l$ is the \emph{loss function}
that quantifies how good the model $w$ explains a data example $d_i \in \mathcal{D}$.

\subsection{Iterative-Convergent ML Algorithms}\label{GD}
A ML job is usually executed by an \emph{iterative-convergent} algorithm
%
%
%
and can be abstracted by the following \emph{additive} operation:
%

$$w^j = w^{j-1} + \alpha\grad(w^{j-1}, S)$$

\noindent
where $w^j$ is the state of the model after iteration $j$
and the update function $\grad$
computes the parameter updates based on some data $S \subseteq \mathcal{D}$
at a learning rate $\alpha$.

Gradient Descent (GD) is arguably the most popular family of iterative-convergent optimization algorithms.
GD is applicable to most of the supervised, semi-supervised, and unsupervised ML problems.
By its name, GD is a class of first-order methods whose update function $\grad$
is based on computing \emph{gradients} from the data.
Batch GD (BGD), Stochastic GD (SGD), Mini Batch GD (MGD), and SVRG++ \cite{ZhuY16icml}
are some example GD family members.
In these algorithms, 
each iteration draws $m$ samples from $\mathcal{D}$ as $S$ 
and the loss of an iteration j is denoted as:
\begin{equation} \label{loss}
 \loss^j = \frac{1}{m}\sum_{i=1}^{m}l(w^j, s_i)
 \end{equation}
where $s_i \in S \subseteq \mathcal{D}$.
Generally, a lower loss indicates a better model accuracy.
Such an algorithm is said to be \emph{converged}
when $w$ stops changing or $\loss^j$ drops below a threshold $\epsilon$.
More specifically, gradient descent learning algorithms theoretically converge
when $\mathbb{E}[R(w^j) - R(w^*)] \le \epsilon$,
where $w^*$ is the optimal model.
Practically, since $w^*$ is unknown, $R(w^j) - R(w^*)$ is approximated by $\loss^j$ to determine convergence.
%
%
Recent works (e.g., \cite{BigModel1, BigModel2, KDD2015}) have shown that many industrial-strength tasks require thousands of iterations or weeks to reach convergence.

%



\subsection{Parameter Server Architecture}
Recently, it is not uncommon to see models with millions of parameters \cite{DeanCMCDLMRSTYN12nips, LiAPSAJLSS14osdi}
and terabytes of training data \cite{tensorflow}.
To support learning at that scale,
the \emph{Parameter Server} (PS) architecture \cite{DeanCMCDLMRSTYN12nips, XingHDKWLZXKY15kdd, LiAPSAJLSS14osdi, MxNet,adam,tensorflow,flexps}
has been advocated to distribute the workload across clusters of nodes.
The concept of a node is abstract.  It can refer to any computation unit like a physical machine, a CPU in a NUMA machine, or a core in a CPU.
Generally, the model parameters are maintained by multiple \emph{server} nodes.
\emph{Worker} nodes periodically ``pull'' (part of) the latest model from the server(s),
perform local computation like calculating stochastic gradient by accessing  their part of training data,
and then ``push'' the updates back to the server(s) whose parameters need to be updated.
Servers update the global model by aggregating the local updates from workers (e.g., averaging the stochastic gradients from workers).
As opposed to traditional message-passing
and MapReduce-like frameworks where pairwise communications between workers are needed in order to exchange each other's parameter updates,
the PS architecture has the advantage of only requiring communications between workers and servers,
thereby mitigating the network bottleneck.
Under the PS architecture, the server-worker ratio is a key knob.

\section{Online Optimization Framework}
\label{tuning}

\begin{figure}
\centering
\includegraphics[width=8cm]{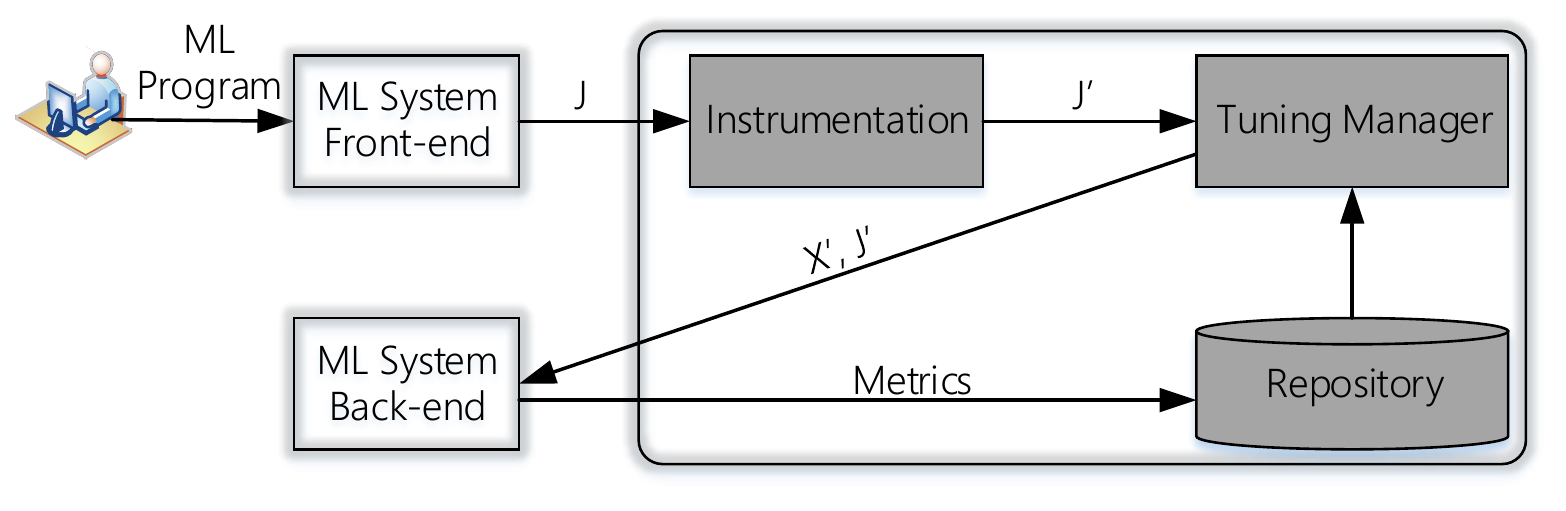}
\caption{Online Optimization Framework}
\label{fig:system}
\end{figure}

%
%
Figure \ref{fig:system} shows our general framework to support self-tuning on PS-style ML systems.
Many ML systems have a front-end and a back-end.
Take {\sf TensorFlow} as an example,
the core operations are carried out by the back-end, which is implemented in C++.
The front-end is responsible for optimizing and orchestrating a job's execution.
The front-end also offers a high-level API for users to write their programs.
There are different {\sf TensorFlow} front-ends but the most popular one is implemented in Python.

On receiving a ML program {\sf J} from the front-end,
the program would be \emph{instrumented}  before sending to the ML system back-end.
The instrumented program would then be executed as a job {\sf J'}, which would emit various  \emph{per-iteration} metrics (e.g., execution time, loss) to a repository during its execution.
%
%
%
Before starting an iteration, the \emph{Tuning Manager} will (1) update a Gaussian Process (GP) model using the metrics collected from the previous iteration,
(2) carry out Bayesian Optimization (BO) to get a possibly new system setting $X'$, and (3) reconfigure the system to setting $X'$.
Practically,  the Tuning Manager would not carry out reconfiguration every iteration
but execute a certain number of iterations for each setting in order to well understand its online statistics efficiency (its live convergence rate).

%
%
%
\subsection{Bayesian Optimization}
Bayesian Optimization (BO) is a strategy for optimizing a  black-box objective function  that is unknown beforehand but  observable through conducting experiments \cite{BOTutorial, BONIPS2012,BAGO}.
Conventionally, 
in BO, each experiment is executed by a different system setting, and the same system setting is used throughout the same experiment. 
%
In our context, each experiment (ML job) is going to be executed by a number of \emph{different} system settings until the job terminates.
%
Therefore, we introduce the loss of the model to the BO's input space so as to 
differentiate whether a system setting is applied to an early stage or to a late stage of the job.
That is important because, in ML, a lousy system setting might improve the loss  when the training just starts,
whereas an optimal setting might hardly improve the loss if the model is converging.

Given an active ML job {\sf J}, the goal of the BO is to recommend the next system setting $X^*$ that is expected to minimize the remaining completion time of {\sf J}.
Let $ X  = \langle c_1 = v_1, \dots, c_d = v_d\rangle$ be a system setting,
where each $c_i$ is a configurable system parameter with value $v_i$.
We use $\T(\langle X, \loss \rangle)$ to denote the remaining completion time of the job
if we switch to setting $X$ 
when the model has reached a loss  $\loss$.
$\langle X, \loss \rangle$ is thus a $(d+1)$-dimensional vector that includes both the system setting and the loss of the model.
%
So, the optimization problem is, given a model whose loss is $\loss'$,
find the $X^*$ that minimizes $T$.
Knowing $\T(\langle X, \loss \rangle)$ ahead of time would be infeasible.
Bayesian Optimization thus returns an approximation solution with little overhead.

We model $\T$ as a GP (Gaussian Process) \cite{gp}
and use BO to suggest the next setting
based on a pre-defined \emph{acquisition function}.
An acquisition function can be updated with more observations.
There are many choices of {acquisition function}
such as (i)  Probability of Improvement (PI) \cite{BONIPS2012}, which picks the next setting that can maximize the probability of improving the current best;
(ii) Expected Improvement (EI) \cite{gp}, which picks the next setting that can maximize the expected improvement over the current best;
(iii) Upper Confidence Bound (UCB) \cite{GPUCB}, which picks the one that has the smallest lower bound in its certainty region.
Different acquisition functions
have different strategies to balance between \emph{exploring} (so that it tends to suggest a possibly new setting from an unknown region of the response surface)
and \emph{exploiting} the  knowledge so far (so that it tends to suggest a setting that lies in a known high performance region).
In this paper, we choose EI because it has shown to be more robust than PI, and unlike UCB, it is parameter-free.
Using BO with EI has the ability to learn the objective function quickly and always return the expected optimal setting.
BO itself is noise resilient.  That is important because
what we can collect from experiments is actually $\T'$:

$$\T' = \T  + e$$

\noindent
where $e$ is a Gaussian noise with zero mean, i.e., $e \sim \mathbb{N}(0, \sigma^2)$.
Since $\T'$, $\T$, and  $e$ are Gaussian,
we can infer $\T$ and its confidence interval~\cite{cherrypick}.
As we discuss in Section \ref{progress} momentarily,
the observation noise comes from the fact $\T'$ is not a direct measurement
but a product between (i) \emph{per-iteration execution time} (hardware efficiency)
and (ii) \emph{estimated number of iterations left} (statistical progress).
Although (i) could be directly observed,
(ii) has to be based on certain empirical estimations (Section \ref{progress}).

BO has an advantage of being {non-parametric},
meaning it does not impose any limit on $\T$, making
our techniques useful for a variety of ML systems.
Furthermore, it
can deal with non-linear response surface but require {far fewer samples} 
than others which have similar power (e.g., deep network).
Lastly, BO has a good track record on tuning database systems \cite{ituned,andyMLtune}.
An approach similar to BO is reinforcement learning \cite{KaelblingLM96} and we will explore that direction as a future work.

\begin{figure}\centering
\scalebox{0.9}{
 \begin{tabular}{ccc}
   \begin{tabular}{c}
$\langle j , X_i, \ya, \yb\rangle$ \\\hline\hline
$\langle 0, X_0, t^0_0, l^0_0 \rangle$ \\
\dots  \\
$\langle 4, X_0, t^4_0, l^4_0 \rangle$ \\
$\langle 5, X_1, t^{5}_1, l^{5}_1 \rangle$ \\
\dots  \\
$\langle 9, X_1, t^{9}_1, l^{9}_1 \rangle$ \\
\dots  \\
$\langle J, X_{b}, t^{J}_s, l^{J}_s \rangle$ \\\hline
   \end{tabular}
         &&
   \begin{tabular}{c}
$\langle X_i, \loss_i, Y_i\rangle$
 \\\hline\hline
$\langle X_0,  l_{init}, Y_0 \rangle$ \\
$\langle X_1,  l^4_0, Y_1 \rangle$ \\
$\langle X_2,  l^{9}_1, Y_2 \rangle$ \\
\dots  \\
$\langle X_{b}, l_{b-1}^{ab-1}, Y_b \rangle$ \\\hline
\multicolumn{1}{l}{
{\footnotesize
$l_{init}$ refers to the loss 
}}\\
\multicolumn{1}{l}{
{\footnotesize
of the initial model
}}\\

\\
   \end{tabular}

    \\
   (a) major execution metrics & & (b) training data
  \end{tabular}
  }
\caption{Major execution metrics and training data}
\label{fig:d}
\end{figure}

\subsection{Initialization Phase}\label{init}
We propose to divide the execution of a ML job into two phases: \emph{initialization} and \emph{online tuning}.
The goal of the intialization phase is to quickly bring in a small set of representative settings and their execution metrics to build the GP.
Initially,
the job starts the first $a$ iterations using the setting $X_0$, which is the default or the one given by the user.
Iterations after that will be executed under $b$ random settings from the setting space,
%
and for each setting it runs $a$ iterations.
Figure \ref{fig:d}a illustrates the major execution metrics that would be inserted into the repository after trying $b$ different settings,
with $a=5$ iterations.
Each record in the collected execution metrics is a quadruple
$\langle j , X_i, \ya, \yb\rangle$,
with
$X_i$ indicates that iteration $j$ was executed using setting $X_i$,
\ya indicates the  execution time of that iteration,
and \yb indicates the loss of the model after that iteration.


\noindent


The loss of \emph{one} iteration alone is insufficient to judge whether a setting has good statistical efficiency.  
Consequently, the execution metrics would be preprocessed into triples
$\langle X_i, \loss_i, Y_i\rangle$,
 where $\loss_i$ is the loss of the iteration just before using setting $X_i$, i.e., $\loss_i = l_{i-1}^{ai-1}$ (e.g.,  $\loss_2 = l^9_1$ in Figure \ref{fig:d}),
 and $Y_i$ is the estimated remaining completion time 
 if starting using $X_i$ from a model with loss $\loss_i$ (details of $Y_i$ in Section \ref{progress}).
Furthermore, it is known that some iterations may incur abnormal loss occasionally \cite{JohnsonZ13nips}.
Therefore, we apply an outliner removal technique in \cite{StatisticalLearning} to remove outliers.
Figure \ref{fig:d}b shows the training data after preprocessing.

%
%

The initialization phase takes a total of $J = a + b\cdot a$ iterations
and it ends with building a GP based on the collected execution metrics (e.g., to compute the parameter values of the kernel function).
After that, the {online tuning} phase starts.

\subsection{Online Tuning Phase}\label{online}
In this phase,
a new setting $X'$ is selected from the GP with the highest expected improvement $EI$
every $a$ iterations.
%
%
%
%
%
%
Depending on the online reconfiguration cost $R_{cost}$ (Section \ref{reconfig}),
if the expected improvement (EI) of $X'$  is larger than $R_{cost}$,
then an online reconfiguration to $X'$ takes place.
In other words, if a reconfiguration costs more than what it will potentially save, that reconfiguration would not take place.
Overall, the online tuning phase goes on until the job finishes.


\subsection{Miscellaneous}
System settings may involve categorical attributes.
For ordinal categorical attributes, we  simply pre-process their values to be integers.
For nominal categorical attributes, we use one-hot encoding \cite{deepLearning} to pre-process their data.
Suppose a categorical attribute contains $k$ categories $\{c_0, c_1, c_2, ..., c_{k-1}\}$, 
in order to denote category $c_i$, 
one-hot encoding represents that value as a $k$-dimensional bit vector with only the $i$-th bit as 1 and all other bits are zero.

We end this section by discussing  how to set the values of $a$ and $b$.
Our major purpose is to avoid our techniques being parametric if possible.
The main usage of $a$ is to deduce the statistical progress (live convergence rate). 
So, we set $a$, the number of iterations executed for each setting, be three times the number of workers so as to assume 
each worker has already pushed the update to the server around three times.
%
%
Our experimental results support our choice empirically.  We regard the theoretical foundation of this choice as a future work.
In this paper, we empirically set $b$, the number of random settings to try in the initialization phase, as 10.
So far, no auto tuners can reach 100\% parameter-free \cite{ituned, andyMLtune}.
%
In Section \ref{conclusion}, we discuss how to possibly eliminate this very last knob, or even the entire initialization phase, by \emph{transfer learning} \cite{tl}.


\section{Online Progress Estimation} \label{progress}

One key input to the online optimization framework is
the estimated remaining completion time $Y_i$ of a job.
Concretely, $Y_i$ can be formulated as:

$$Y_i = \bar{t_i} \times r_i^j$$

\noindent
which is a product between (i) \emph{per-iteration execution time} $\bar{t_i}$ (hardware efficiency)
and (ii) \emph{estimated number of iterations left} $r^j_i$ (statistical progress).
$\bar{t_i}$ could be directly computed as the average of the recorded iteration times of using that setting $X_i$, e.g., $\bar{t_1}$ can be computed as $(t^5_1+ t^6_1 + \cdots + t^9_1)/5$ in Figure~\ref{fig:d}a.

In this section, we focus on $r_i^j$, the remaining number of iterations required to reach model convergence.
We posit that 
estimating $r_i^j$ in ML systems is as challenging
as estimating cardinalities in query optimization.
It is known that today we could still find errors up to orders of magnitude in cardinality estimation techniques \cite{HarmouchN17}.
Yet, most query optimizers live with that in practice.
So following decades of experiences from query optimization \cite{QOSolved},
we aim for estimates that would not lead us to disastrous  settings, instead of perfect estimates that are not demanded in practice.

\subsection{From Bounds to Legitimate Estimation } \label{online2}

Studying the convergence rate of various learning algorithms is a very active research topic in machine learning
\cite{meng2017convergence,RechtRWN11nips,shalev2014understanding,nemirovski2009robust}.
Since parameter server is a parallel learning architecture,
in this paper we focus on \emph{parallel} gradient descent learning algorithms.
For example, Hogwild! \cite{RechtRWN11nips} is a parallel 
 gradient descent learning algorithm that would theoretically converge,
 i.e., $\mathbb{E}[R(w^z) - R(w^*)] \le \epsilon$
\emph{after} $z$
iterations, where

\begin{equation}\label{bound}
z \ge \frac{\mathcal{H}}{\epsilon}\log\frac{d}{\epsilon}
\end{equation}

\noindent
under the assumptions of $R$ is c-strongly convex and L-smooth,
and $\mathcal{H}$ is the hidden constant, $d = L||w^{0} - w^*||^2$ and $L$ is the Lipschiz constant,
$w^*$ is the optimal model parameter, 
$w^0$ is the initial model parameter, and $\epsilon$ is the user-specific convergence threshold.

%
%
%


%
%



In this paper, we use Hogwild! as an example 
(because its convergence results apply to both bulk synchronous and asynchronous model update)
and generalize its offline convergence analysis to be an online estimation function.
Without loss of generality, 
assume the learning has finished $j_0$ iterations
and switches to use setting $X_i$ to execute $a$ more iterations.
%
%
%
%
Then, we know the $a$ new pairs of 
$\{\langle j, \yb \rangle \}_{j=j_0+1}^{j=j_0+a}$
would be scattered around the curve of:

\begin{equation}\label{fit2}
j = j_0 + \frac{\mathcal{H}_i}{\yb}\log\frac{d_i}{\yb} 
\end{equation}

Fitting those $a$ pairs of $\langle j, \yb \rangle$ to Equation \ref{fit2} 
could then determine the values of $\mathcal{H}_i$ and $d_i$ for setting $X_i$.
When $\mathcal{H}_i$ and $d_i$ are known, we can estimate $r_i^j$ as:

\begin{equation}\label{r}
r_i^j =  \frac{\mathcal{H}_i}{\epsilon}\log\frac{d_i}{\epsilon}  
\end{equation}

This methodology was pioneered by \cite{gdoptimizer}, but focused on gradient based algorithms under 
a  \emph{serial} and \emph{offline} setting.
For example, they deduce the total number of  iterations as
 $z = \mathcal{H}/\epsilon$, with only one hidden constant.
For \emph{online} tuning, the fitting needs to be carried out multiple times,
once for each different setting $X_i$ from different starting point $j_0$ (c.f. Equation \ref{fit2}).
Furthermore, for gradient based algorithms 
under a \emph{parallel}  setting, like Hogwild!, their convergence guarantees 
usually capture both the algorithmic factor $d_i$ (e.g., use of risk function) 
and the environmental factor $\mathcal{H}_i$ (e.g., data distribution, network delay).
While  $\mathcal{H}_i$ is explicitly related to $\langle j, \yb \rangle$,
the relationship between the algorithmic factor $d_i$ and live metric $\langle j, \yb \rangle$ is implicit.  
To determine $d_i$, we first derive its upper bound.
Specifically, we show that $d_i \le 2q\loss^{j}$:

\begin{enumerate}[(i)]
\item $\loss^{j} = R(w^{j}) - R(w^*)$  \verb|     | //Section \ref{GD}
\item $d_i = L||w^{j} - w^*||^2$ 
\item $||w^{j} - w^*||^2 \le \frac{2}{c}(R(w^{j}) - R(w^*))$ \verb|| // $R(w)$ is c-strongly convex
\item $||w^{j} - w^*||^2 \ge \frac{2}{L}(R(w^{j}) - R(w^*))$ \verb|	| // $R(w)$ is L-smooth

\item Combine (i), (ii) and (iii), we have $d_i \le \frac{2L}{c}\loss^{j}$

\item Combine (iii) and (iv), we get $\frac{L}{c} \ge 1$

\item Consider a constant $q = \frac{L}{c} \ge 1$, then we have $d_i \le 2q\loss^{j}$

\end{enumerate}

Now the question becomes how to determine $d_i$ (and $\mathcal{H}_i$) using 
the $a$ collected $\langle j, \yb \rangle$ pairs.
Notice that we should not fit both $\mathcal{H}_i$ and $d_i$ together because that may find a value for $d_i$ exceeding its upper bound.
But up to this point, we understand that (a) $d_i$ should not be a large number,
because of the log term in Equation \ref{fit2} means $\log d_i - \log \yb$
and the difference between $\yb$ and $l^{j+1}_i$ is tiny in practice.
As such, a large $d_i$ would dominate the term  $\log\frac{d_i}{l^j_i}$ and degenerate it to a constant.
%
%
On the other hand, we understand (b) if $d_i$ is smaller than many of the $\yb$'s  collected, 
then during fitting many $\log\frac{d_i}{l^{j}_i}$'s would result in negatives, making $\mathcal{H}_i$ be fitted as negative.
That would make Equation \ref{r} returns negative numbers as results, which is undesirable.
Based on these information, we set $d_i$ as 

\begin{equation}\label{d}
\min\{2l_i^{j_0}, \max\{l_i^{j_0+1}, l^{j_0+2}_i, \cdots, l^{j_0+a}_i\}\}
\end{equation}
\noindent
and then deduce $\mathcal{H}_i$ based on the $a$ collected $\langle j, \yb \rangle$ pairs.
In Equation \ref{d}, the term $2l_i^{j_0}$ is the supremum of $d_i$ based on its upper bound to address concern (a)
and the $\max$ term inside addresses concern (b).

%
%

\subsection{ Limitations and Opportunities} \label{limit}


The foundation of our statistical progress estimation (and also \cite{gdoptimizer}) are based on known theoretical convergence bounds of the various learning algorithms.
The advantage of this approach is that in principle anyone can follow our methodology to 
improve the estimation function whenever there are new results on the bounds (e.g., tighter bounds that consider more factors).
Nonetheless,
while the convergence bounds of many ML problems are known,
the convergence bounds of a number of non-convex problems are still under development.
For example, there are bounds for non-convex PCA \cite{NonConvexOptimization}
and two-layer neural networks with ReLU \cite{ConvergenceAnalysis},
but bounds for deeper neural networks with many layers are still being developed.
For our {\sf \name} prototype, we have implemented an estimation function based on Hogwild! (which assumes problems are convex).
Our experiments show that our estimation function is able to avoid disastrous settings 
and return efficient system settings for two convex and one non-convex problems.
We regard this system paper as an initial effort and 
we will develop more specific estimation functions for each class of ML models and learning algorithms with known convergence bounds.
This scale of work would however require effort of the community and 
we open-source our prototype to facilitate that.

\section{Online Reconfiguration}
\label{reconfig}

Online reconfiguration changes the system setting in the course of a ML job.
Under the PS architecture, the following physical changes could be triggered by a reconfiguration:

\begin{itemize}
\item (Type I) Data Relocation: 
For example, a recommendation that suggest turning a worker node to a server node would trigger this type of reconfiguration.
Here, we further bifurcate data relocation into:
\begin{itemize}
\item (Type I-a) Training Data Relocation
\item (Type I-b) Model Data Relocation
\end{itemize}
\item (Type II) System Setting Reconfiguration:   For example, in {\sf TensorFlow}, there is a knob to turn on or off the function inlining optimization.  This kind of knobs would not trigger any data relocation.
\end{itemize}

To implement online reconfiguration,
a baseline solution is to re-use the system's checkpointing and recovery feature (e.g., the save \& restore in {\sf TensorFlow}).
In most circumstances, that feature is collectively implemented by four techniques: \\

\begin{enumerate}
\item{Checkpointing (CKP)}:~~~This saves the model state (e.g, the current model $\w^j$, the current iteration number $j$) to a persistent storage.
Usually, this would not save any system settings (e.g., whether function inlining is on or off) because those values are stored separately in a system configuration/property file/in-memory data structure.
Moreover, checkpointing does not involve the training data because there is a master copy of the training data in the shared storage (e.g., HDFS).
\item{System Setting Recovery (SSR)}:~~~This is built-in as part of the recovery process, in which the system is reinitialized based on the setting specified in the configuration/property file/data structure.
\item{Model Data Recovery (MDR)}:~~~This is the other part of the build-in recovery process, in which the model state is restored to the servers based on the system setting.
\item{Training Data Recovery (TDR)}:~~~Because the training data is read only and stored in the shared storage.
Therefore, on recovery, the workers would simply fetch the missing data from the shared storage directly.
\end{enumerate}

Existing ML systems implement their checkpointing and recovery process as a CKP and a full SSR+MDR+TDR, respectively.
We regard that as the baseline reconfiguration implementation.
It is expected this baseline implementation incurs high overhead (e.g., checkpointing the state) and cause system quiescence. 
In this paper, we have 
developed a new scheme that can carry out online reconfiguration more efficiently.
Before introducing that, we first present a new technique for carrying out Type I-b reconfiguration efficiently.

\stitle{On-Demand-Model-Relocation (ODMR)}~~~In our experience of applying our techniques on {\sf TensorFlow}, a lion's share of reconfiguration cost attributes to Type I-b, i.e.,
 the cost of relocating some model parameters from one node to another node (e.g., when a recommendation suggests increasing the number of servers).
Consequently, 
we design a technique, namely, On-Demand-Model-Relocation, that can carry out more efficient Type I-b model data relocation.

The idea of ODMR is to carry out parameter relocation on demand.
Concretely, on receiving a Type I-b request, the system only invokes SSR to reflect the decision of moving a parameter from a source to a destination.
The actual parameter movement takes place only when a parameter is pulled from the source server and pushes back to the destination server.
Suppose there are two servers $S_1$ and $S_2$ and they originally manage parameters $\{w_1, \dots, w_6\}$ and $\{w_7, \dots, w_{12}\}$, respectively.
Now, assume a reconfiguration suggests to add one more server $S_3$ so that the three servers, $S_1$, $S_2$, and $S_3$ manage parameters
$\{w_1, \dots, w_4\}$, $\{w_7, \dots, w_{10}\}$, and $\{w_5, w_6, w_{11}, w_{12}\}$, respectively.
So, when a worker requests a parameter that is supposed to be relocated, e.g., $w_{12}$,
we simply let the worker to pull from the old destination $S_2$.  After the workers have computed the updates,
they push both their original values and the updates to the new destination $S_3$.
The reasons of pushing the original value are that (1) the destination $S_3$ does not have the original value $o$, so sending the updates $u$ alone is not enough and (2) the original value ``flags'' the servers that this push is special and to avoid possibly repeated counting --- the first time the server receives the message $\langle o,u_1\rangle$ it should create a new parameter with value $o+u_1$, but the second time it receives a message $\langle o,u_2\rangle$, it should act like receiving  a normal push with $u_2$.

The ODMR approach has the merit of overlapping a Type I-b relocation with the usual push-and-pull operations.
It would not cause any system quiescence as the basic solution does.
By mix-and-match the existing checkpointing and recovery techniques in PS-style systems and ODMR,
our reconfiguration scheme is as follows:

\begin{itemize}
\item For Type I-a reconfiguration only,   invoke TDR.
\item For Type I-b reconfiguration only,  invoke ODMR.
\item For Type II reconfiguration only, change the system configuration file and invoke SSR.
\item For any combination of the above, invoke the union of their actions.
\end{itemize}

%
%

We end this section by recalling the need to estimate the reconfiguration cost $R_{cost}$ for the online tuning phase (Section \ref{online}).
With the discussion above, it becomes clear that $R_{cost}$ depends on the reconfiguration type and technique.
Nonetheless, empirically we observe that the cost variance of each technique  is small
and thus we can simply deduce the costs of each individual techniques from the execution metrics collected during the initialization phase.

\section{Experiments} \label{exp}


We implemented our techniques on top of {\sf TensorFlow} v1.8.
The details about our implementation can be found in Appendix~\ref{app:B}.
Briefly, the implementation includes a user-level library written in Python 2.7 in order to abstract out the system setting of a {\sf TensorFlow} program.
We implemented the Tuning Manager and the repository using Python.
We modified the front-end  of  {\sf TensorFlow} so as to support our reconfiguration scheme.
We refer this prototype implementation as {\sf \name} in this section.
{\sf \name} can support both asynchronous parallel (ASP) and bulk synchronous parallel (BSP) training.
Table \ref{knobs} lists all the system knobs supported by  {\sf \name}.

\begin{table*} 
    \caption{System knobs tuned in {\sf \name}}\label{knobs}
\centering
\scalebox{0.9}{\small
    \begin{tabular}{|p{8cm}|p{10cm}|}
    \hline
    Knob  & Meaning \\ \hline\hline
\emph{tf.ClusterSpec::ps} & The number of parameter servers \\
\emph{tf.ClusterSpec::worker} & The number of workers \\
\emph{tf.ConfigProto::intra\_op\_parallelism\_threads} & The number of thread of thread pool for the execution of an individual operation that can be parallelized \\
\emph{tf.ConfigProto::inter\_op\_parallelism\_threads} & The number of thread of thread pool for operations that perform blocking operations \\
\emph{tf.OptimizerOptions::do\_common\_subexpression\_elimination}  & A switch to enable common subexpression elimination  \\
\emph{tf.OptimizerOptions::max\_folded\_constant\_in\_bytes} & The total size of tensor that can be replaced by constant folding optimization \\
\emph{tf.OptimizerOptions::do\_function\_inlining} & A switch to enable function inlining \\
\emph{tf.OptimizerOptions::global\_jit\_level} & The optimization level of jit compiler: \{OFF, ON\_1, ON\_2\} \\
\emph{tf.GraphOptions::infer\_shapes} & Annotate each Tensor with Tensorflow Operation output shape \\
\emph{tf.GraphOptions::place\_pruned\_graph} & A switch to place the subgraphs that are run only, rather than the entire graph \\
\emph{tf.GraphOptions::enable\_bfloat16\_sendrecv} & A switch to transfer float values between processes using 16 bit float \\
    \hline
    \end{tabular}}

\end{table*}

\stitle{Hardware} We performed all the experiments on a cluster of 36 identical  servers, connected by Ethernet.
The network bandwidth is 10Gbps.
The computing nodes run 64-bit CentOS 7.3, with the training datasets on HDFS 2.6.0.
Each node is Intel Xeon E5-2620v4 system with 16 cores CPU running at 2.1 GHz,  64GB of memory, and 800GB SSD.

\stitle{Comparison}  For comparison purposes, we 
use vanilla {\sf TensorFlow} as the baseline and compare  it with {\sf \name}.
For each experiment executed by {\sf TensorFlow}, 
we repeated the job 100 times, each using a different \emph{random system setting} and report:

\begin{enumerate}
  \item {\sf Worst}: the worst completion time of {\sf TensorFlow} among 100 random settings.
  \item {\sf Average}: the average completion time of {\sf TensorFlow} among 100 random settings.
  \item {\sf Best}: the best completion time of {\sf TensorFlow} among 100 random settings.
\end{enumerate}

Note that the result of {\sf Best} could not be achievable in practice.
That is because nobody would really run the same job on {\sf TensorFlow} a hundred times just for identifying the best system setting for a particular model (hyperparameter) and dataset.
We also note that {\sf TensorFlow}, although popular, is more like a software library than a system --- currently users must explicitly specify most system parameters (e.g., the number of workers)
and there are {\bf no} default values per se.  

\begin{table}
    \caption{Training Datasets and Workloads}\label{datasets}
    \centering
    \footnotesize
    \scalebox{0.85}{
	    \begin{tabular}{|c|c||c|c||c|}
	    \hline
	    Dataset & Data Size & Model & Model Size  & Model Update\\ 
	    &    &   & (\# of parameters) &  \\ \hline\hline
	    KDD12 \cite{KDD12} & 17G & LogR  & 109,372,904  & Asynchronous \\
	    CRITEO \cite{criteo} & 2.2T & SVM  & 1,000,000 & Asynchronous \\
	    CIFAR10 ~\cite{data1} & 160M & CNN  & 654,968  & Asynchronous\\ 
	    ImageNet8 ~\cite{Om2016} & 1.2G & CNN & 58,289,352  &Bulk Synchronous \\ \hline
	    \end{tabular}}
\end{table}

\stitle{Workload and Datasets}
We evaluate {\sf \name} with three widely used machine learning models:
(i) $l_2$ regularized Logistic Regression (LogR),
(ii) Support Vector Machine (SVM), and
(iii) Convolutional Neural Network (CNN).
For CNN, we used AlexNet \cite{alexnet}, a convolutional neural network with five layers.
AlexNet is also used in many work \cite{lecun1998gradient,adam}.
We remark that AlexNet is non-convex, and its convergence bound is still actively researched by the machine learning community.
But as a system paper, we still try AlexNet to see if we can use our current estimation function as a heuristic.
This is not uncommon in machine learning.  For example, it is known that SGD might converge to a saddle point / local optimal but not global optimum when facing non-convex problems.
Nonetheless, SGD is still being extensively used in all sort of deep learning problems in practice.

Table~\ref{datasets} summarizes the characteristics of the datasets and the workloads used in our experiments.
For CNN, we used two different datasets.
Specifically, ImageNet is a typical dataset for deep learning.
However, each training job on ImageNet can take weeks on our cluster.  When running {\sf TensorFlow} 
in the baseline experiments, some random (poor) settings took even a longer time to finish.
As we have to run a lot of baseline experiments,
we follow \cite{Om2016} to use a reduced version of ImageNet, namely ImageNet8, which consists of the first 8 classes in the original data.
The convergence thresholds $\epsilon$ for LogR, SVM, and CNN are  set as 0.2, 0.98, 0.5, and 1.5, respectively.
These thresholds are chosen to ensure we can obtain the baseline results within months.




\begin{figure*}
	\begin{center}
		\subfigure[LogR]{\label{fig:completion-time-lr}
			\includegraphics[width=0.47\columnwidth]{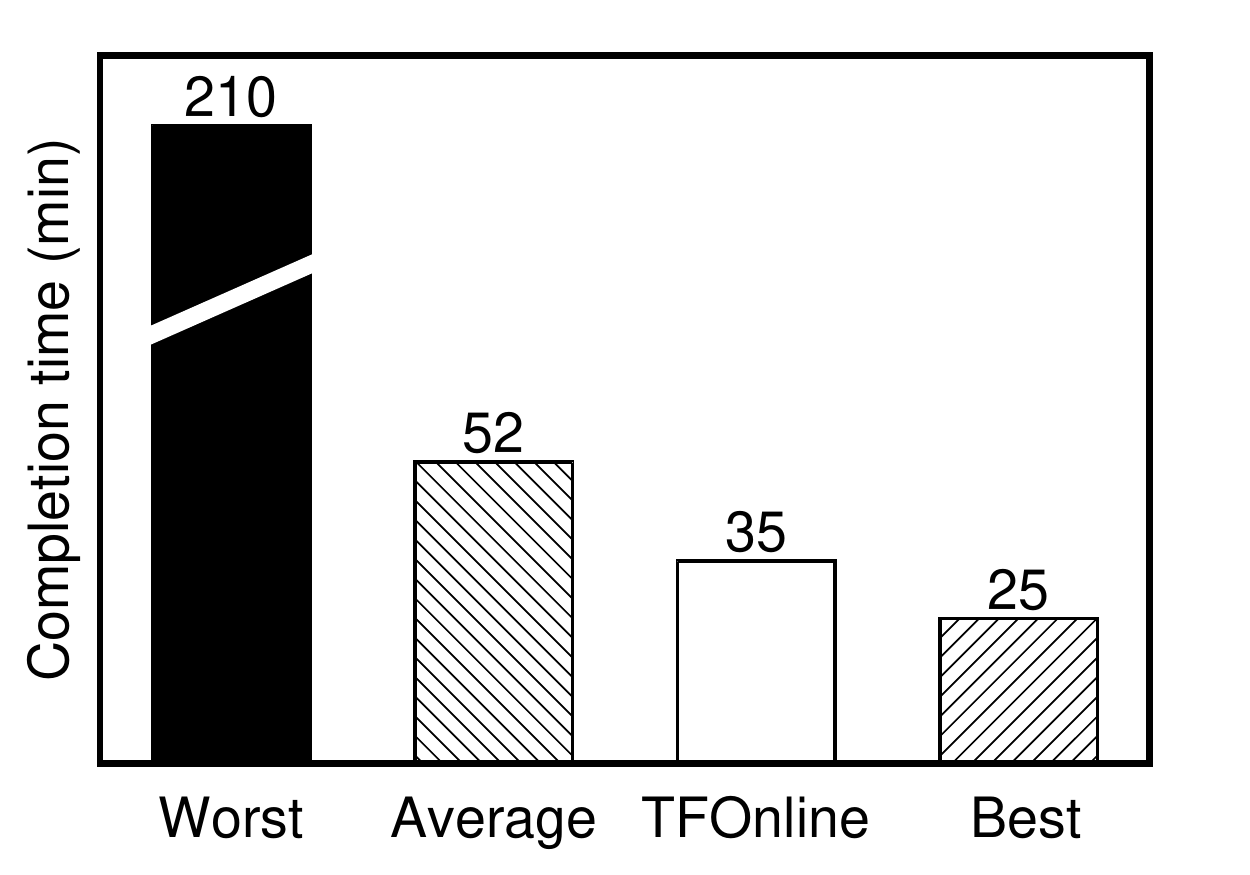}
		}
		\subfigure[SVM]{\label{fig:completion-time-svm}
			\includegraphics[width=0.47\columnwidth]{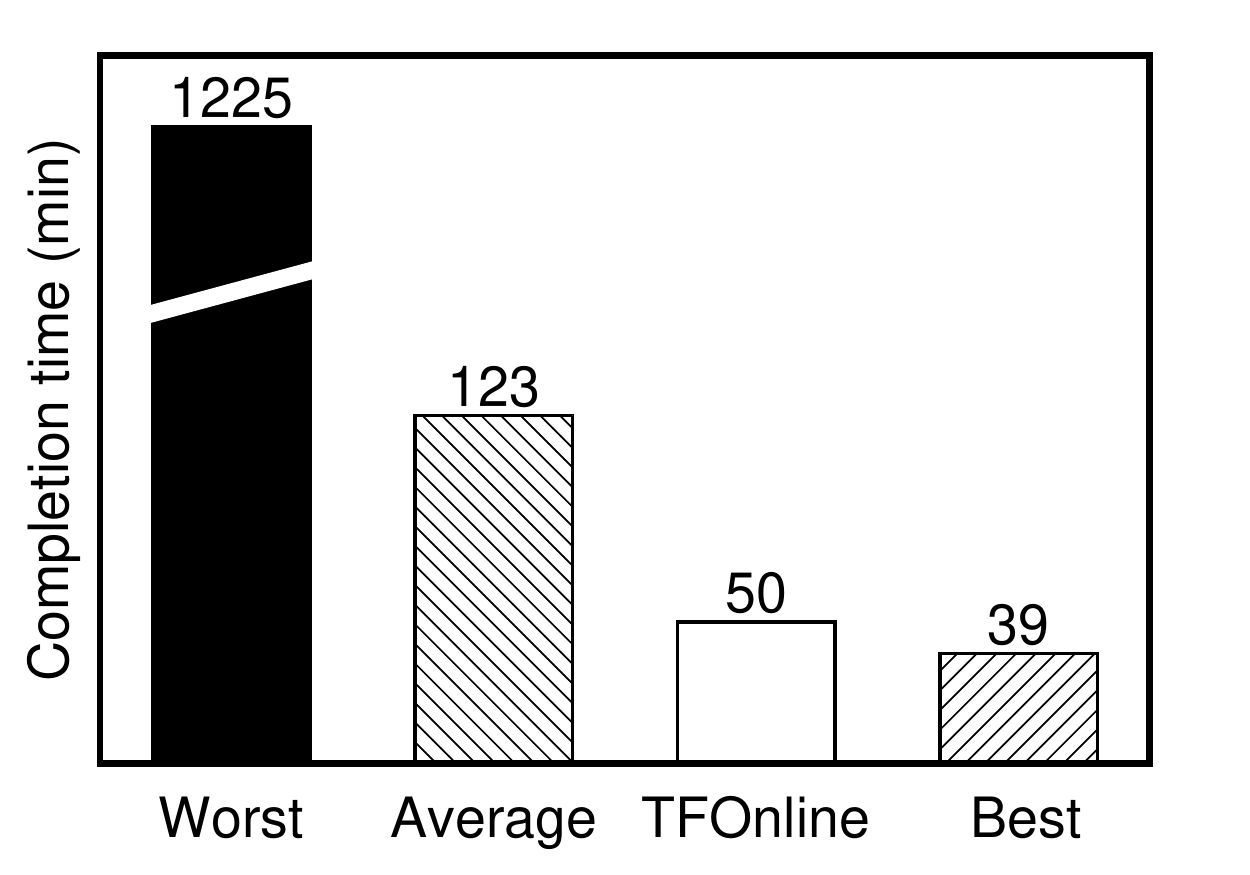}
		}
		\subfigure[CNN on CIFAR]{\label{fig:completion-time-nn}
			\includegraphics[width=0.47\columnwidth]{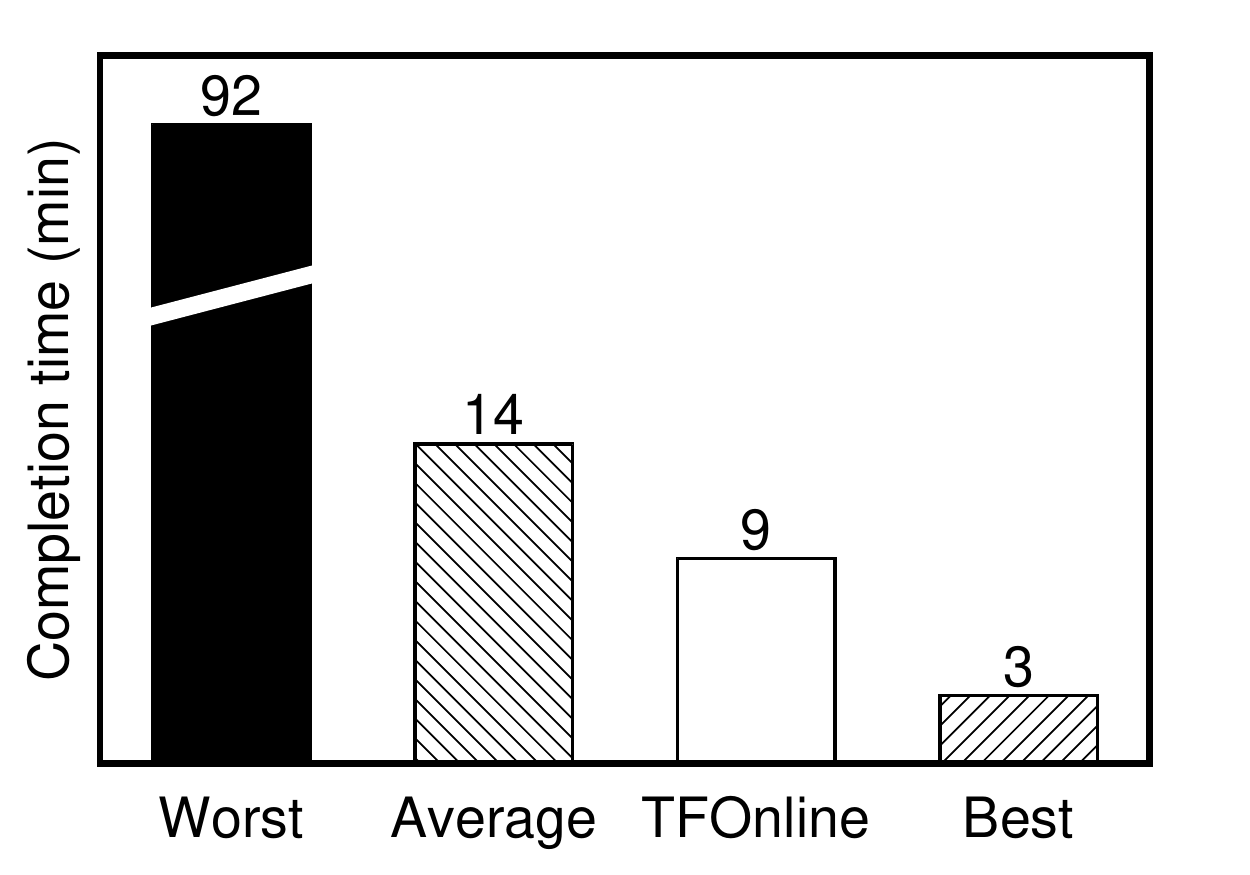}
		}
		\subfigure[CNN on Imagenet8]{\label{fig:completion-time-imagenet}
			\includegraphics[width=0.47\columnwidth]{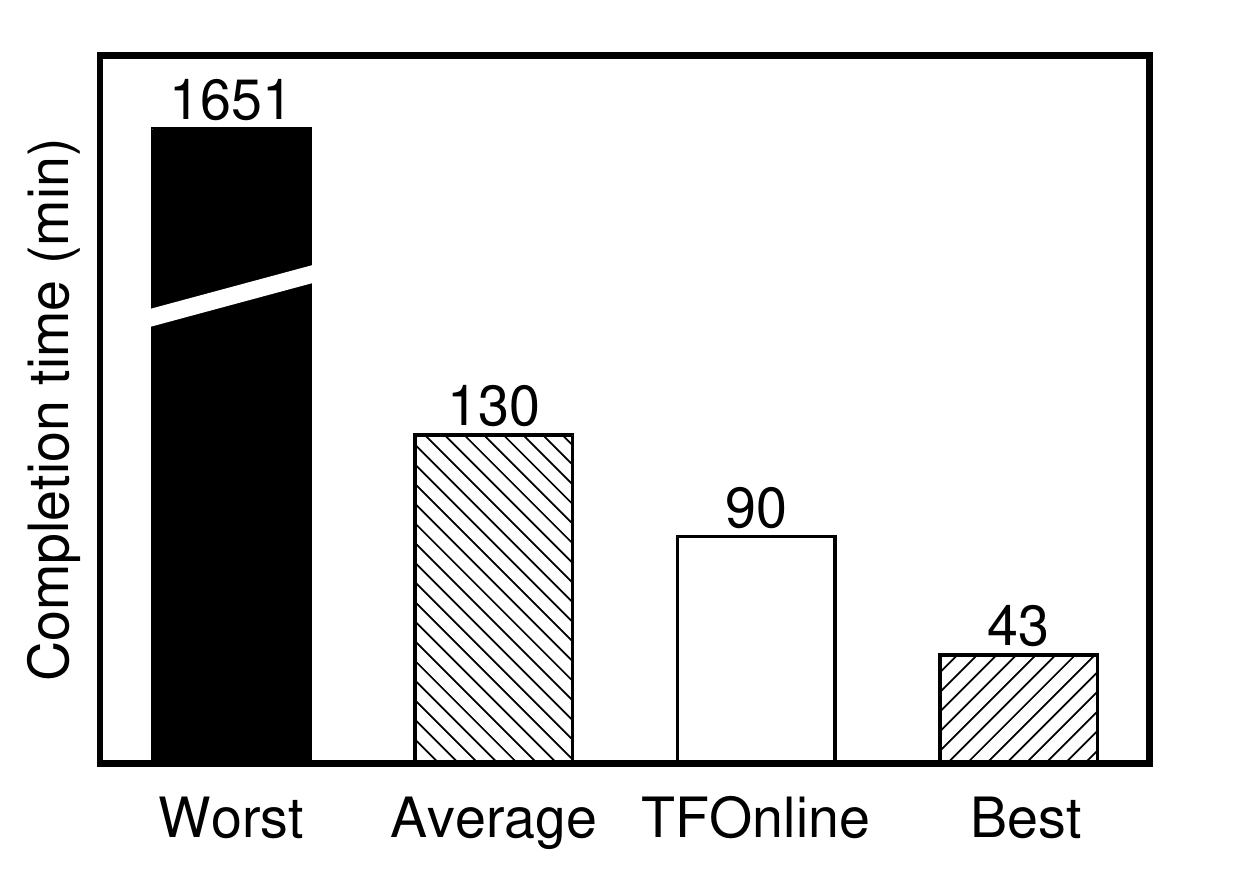}
		}
		\caption{End-to-end completion time comparison}\label{fig:completion-time}
	\end{center}
\end{figure*}

\begin{figure*}
\begin{center}
	\subfigure[LogR]{\label{fig:loss-time-lr}
	\includegraphics[width=0.47\columnwidth]{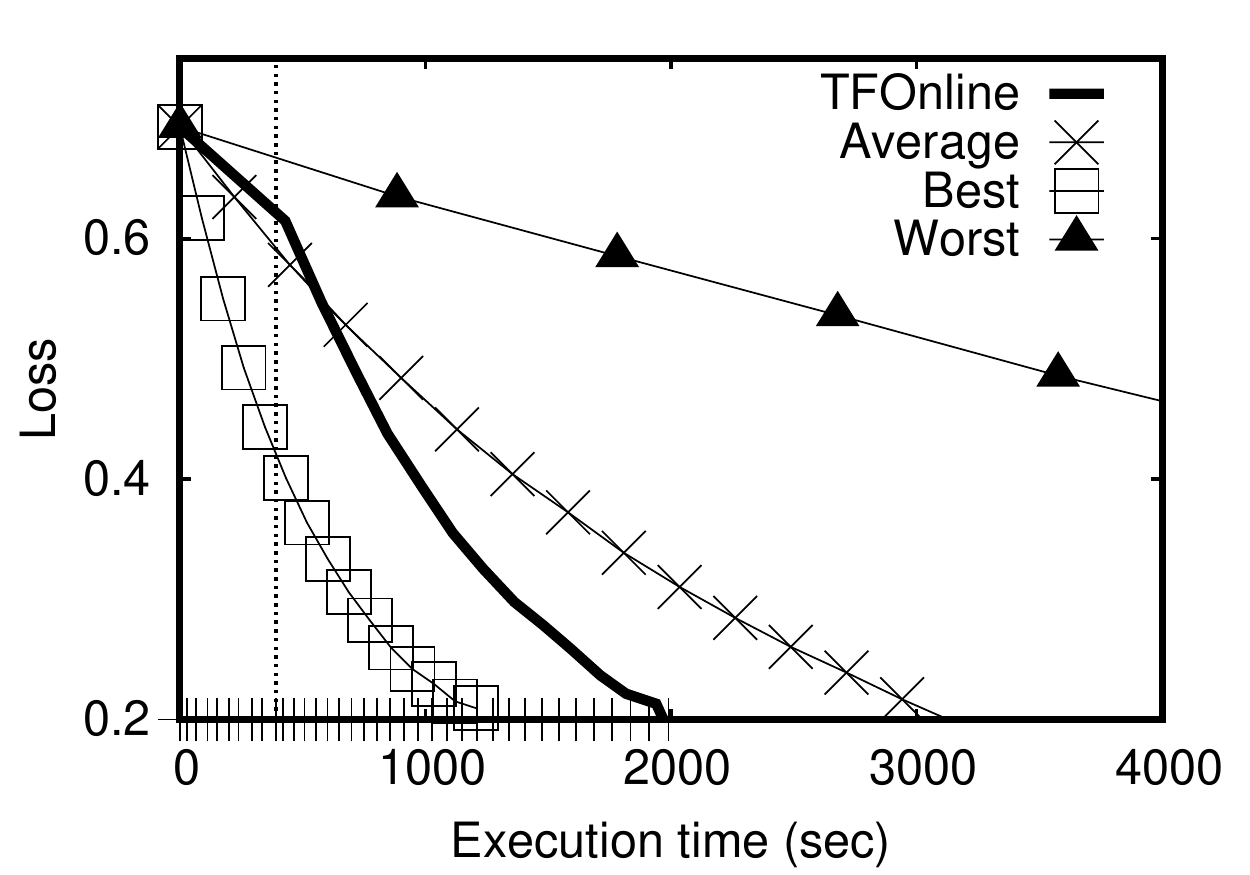}
    }
	\subfigure[SVM]{\label{fig:loss-time-svm}
	\includegraphics[width=0.47\columnwidth]{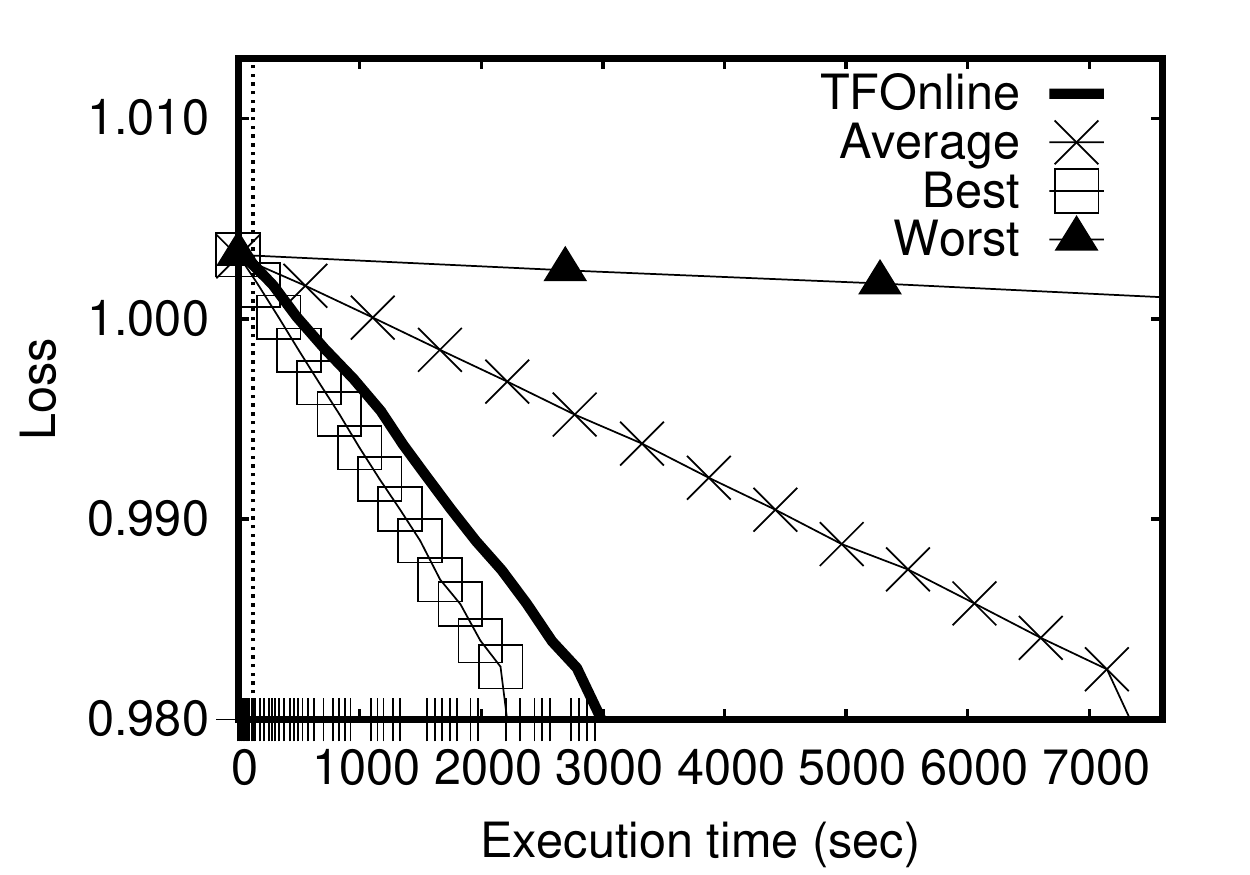}
    }
   \subfigure[CNN on CIFAR]{\label{fig:loss-time-nn}
	\includegraphics[width=0.47\columnwidth]{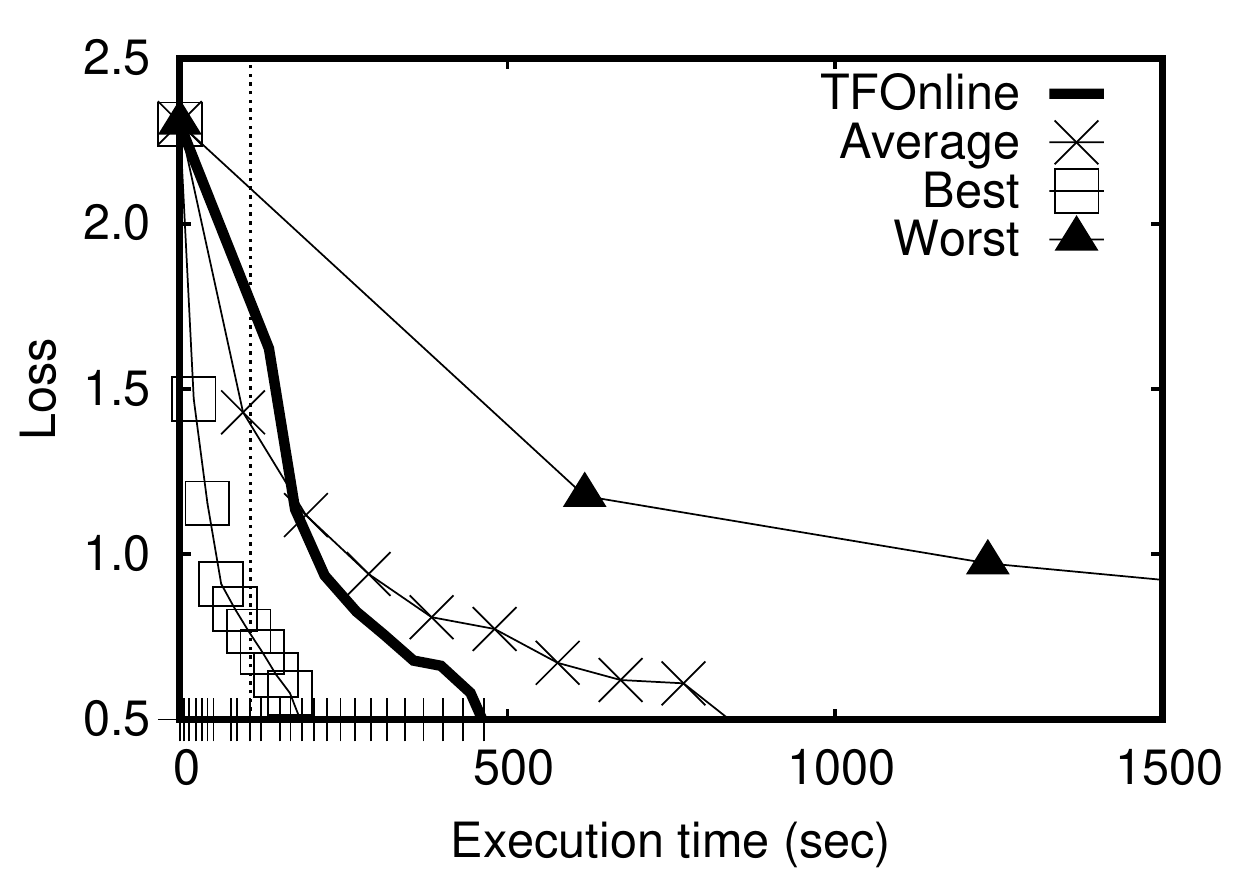}
    }
	\subfigure[CNN on ImageNet8]{\label{fig:loss-time-alexnet}
	  \includegraphics[width=0.47\columnwidth]{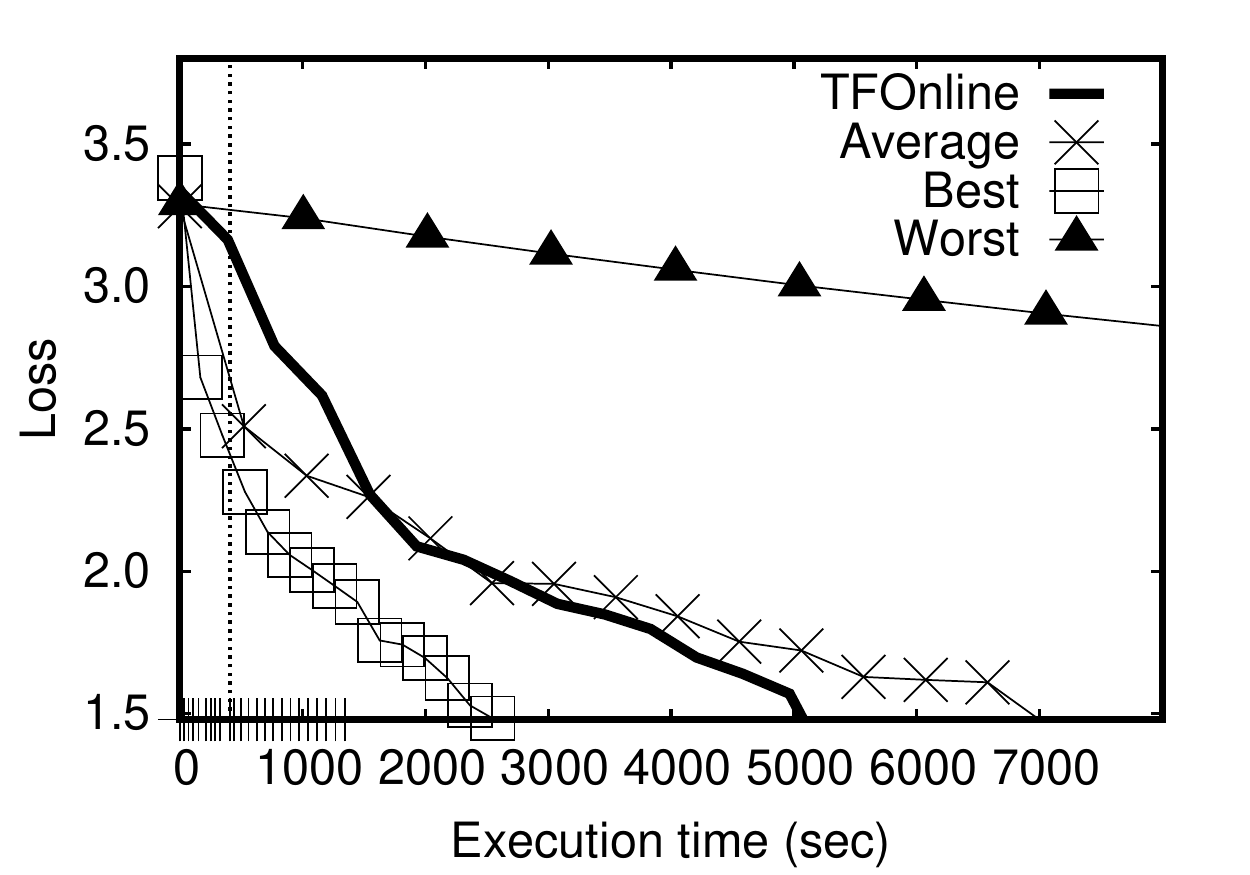}
    }
\caption{Loss vs. Job training time} \label{fig:loss-time}
\end{center}
\end{figure*}

\subsection{Performance Evaluation}
%
%
%

Figure~\ref{fig:completion-time} compares the completion time of {\sf \name} with {\sf TensorFlow}.
We see that {\sf \name} has about 1.4$\times$ (CNN on ImageNet8) to 2.5$\times$ (SVM) speedup when compared with {\sf Average}, 
meaning {\sf \name} saves much time for average ML users who have little system background.
Furthermore, {\sf \name} helps users to avoid disastrous bad settings,
which are 6$\times$ (LogR) to 18$\times$ (CNN on ImageNet8) slower than using {\sf \name}.

%
%

Figure~\ref{fig:loss-time} shows that the loss of the jobs with respect to the job training time.
In the figure, we indicate the moment when {\sf \name} switches from its initialization phase to the online tuning phase with a vertical dotted line.
We also add a marker on the x-axis whenever {\sf \name} changes the system setting online.
From the figure, we observe that {\sf \name} might have a slower convergence rate during the initialization phase 
because it was trying different settings and some of those might not be good ones.
Nonetheless, we know that is worth doing because once {\sf \name} enters the online tuning phase, it progressively uses better system settings and converges much faster afterwards.

\begin{figure*}
\begin{center}
	\subfigure[LogR]{\label{fig:loss-iter-lr}
	\includegraphics[width=0.47\columnwidth]{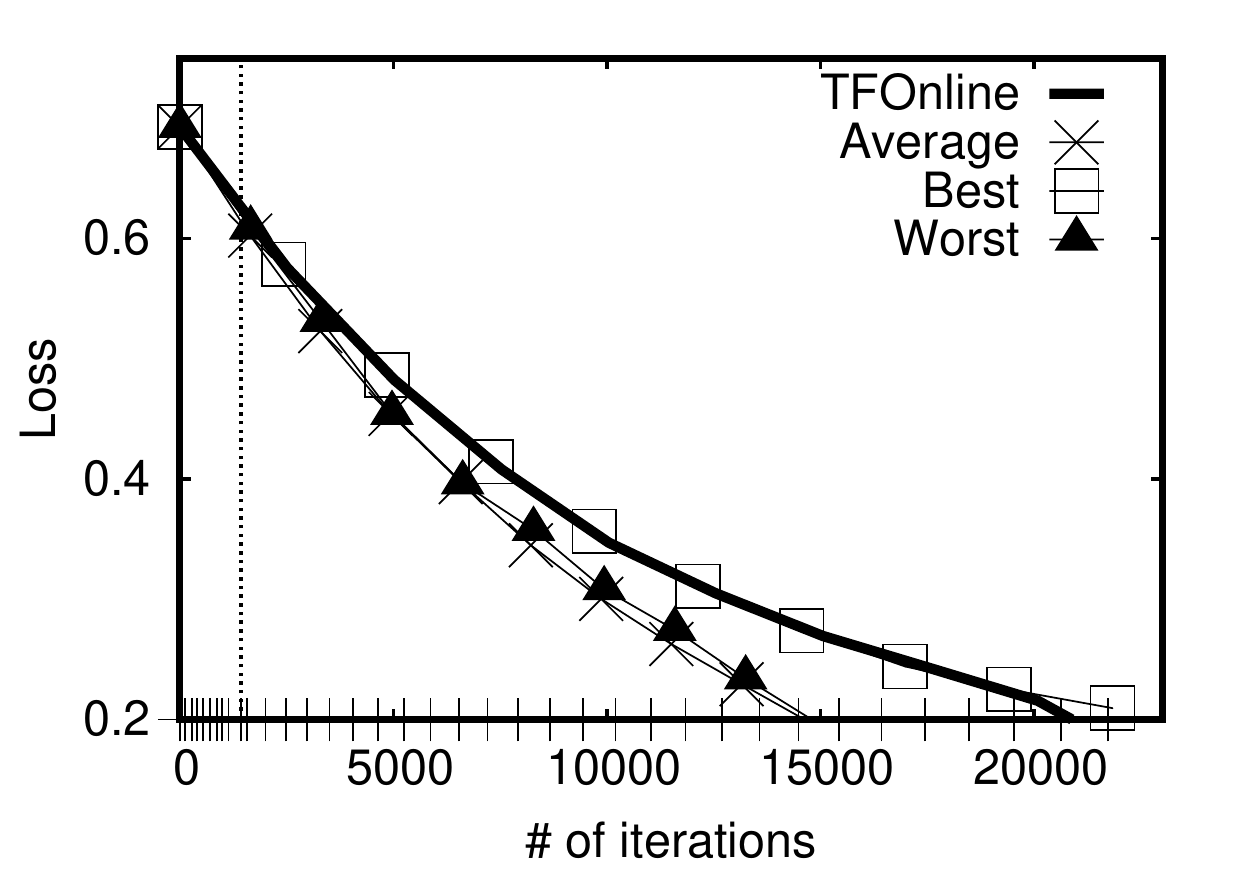}
    }
	\subfigure[SVM]{\label{fig:loss-iter-svm}
	\includegraphics[width=0.47\columnwidth]{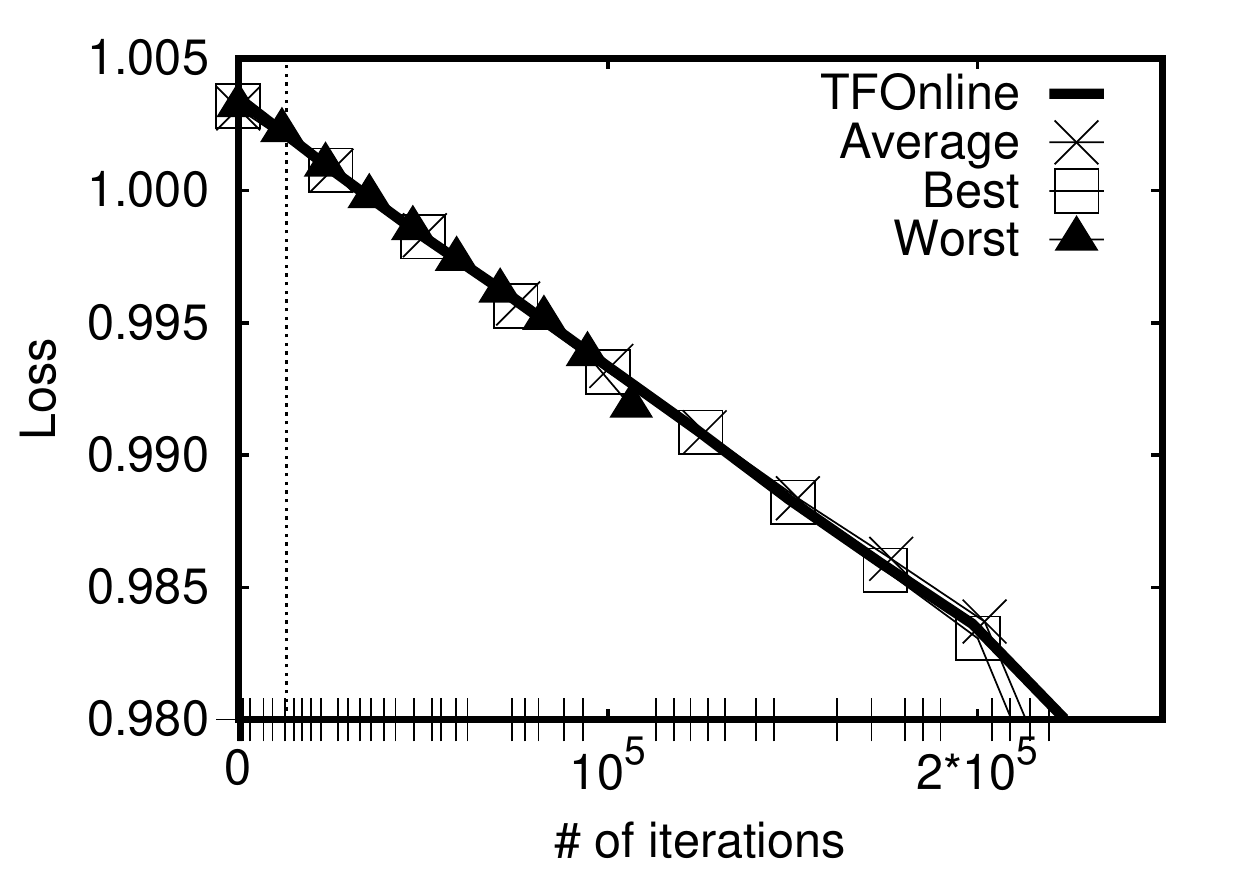}
    }
	\subfigure[CNN on CIFAR]{\label{fig:loss-iter-nn}
	\includegraphics[width=0.47\columnwidth]{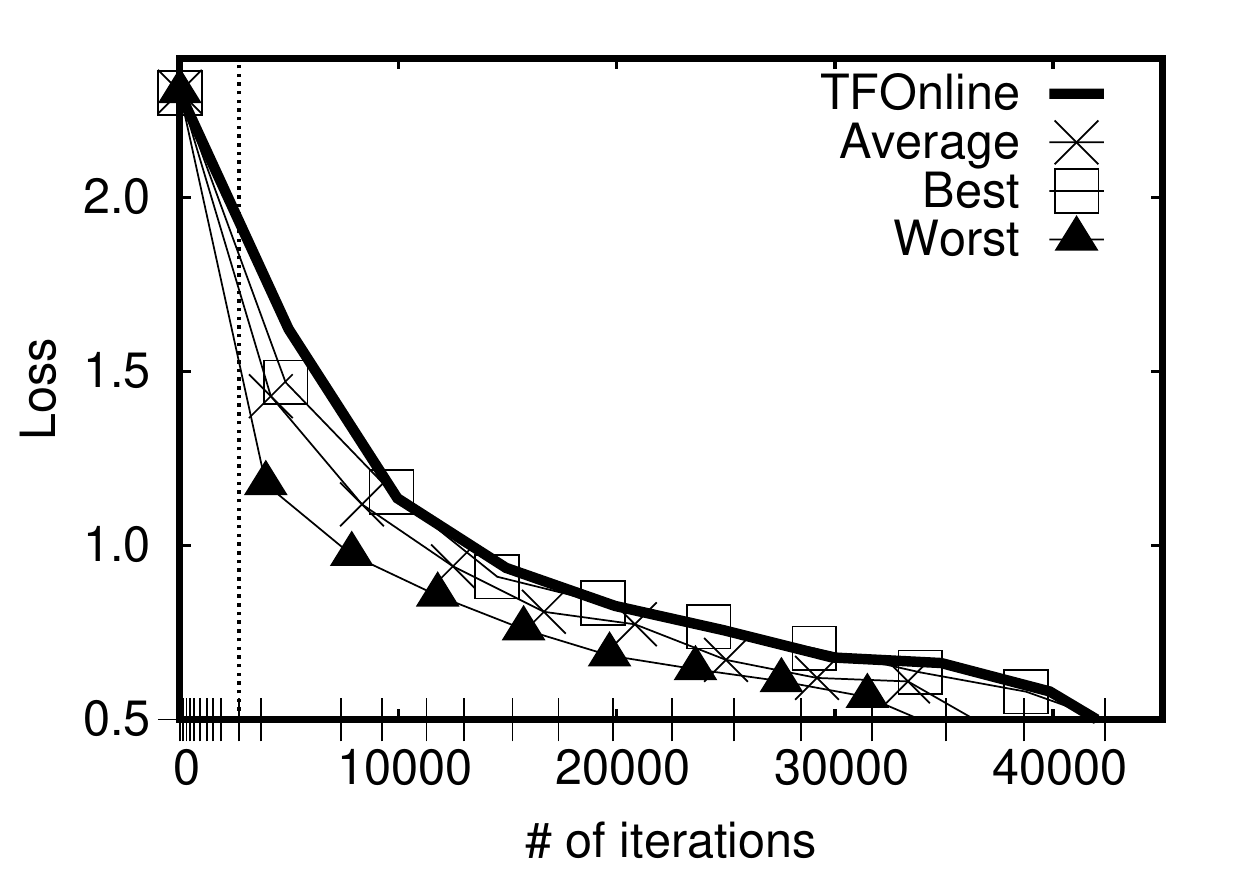}
    }
	\subfigure[CNN on ImageNet8]{\label{fig:loss-iter-alexnet}
	\includegraphics[width=0.47\columnwidth]{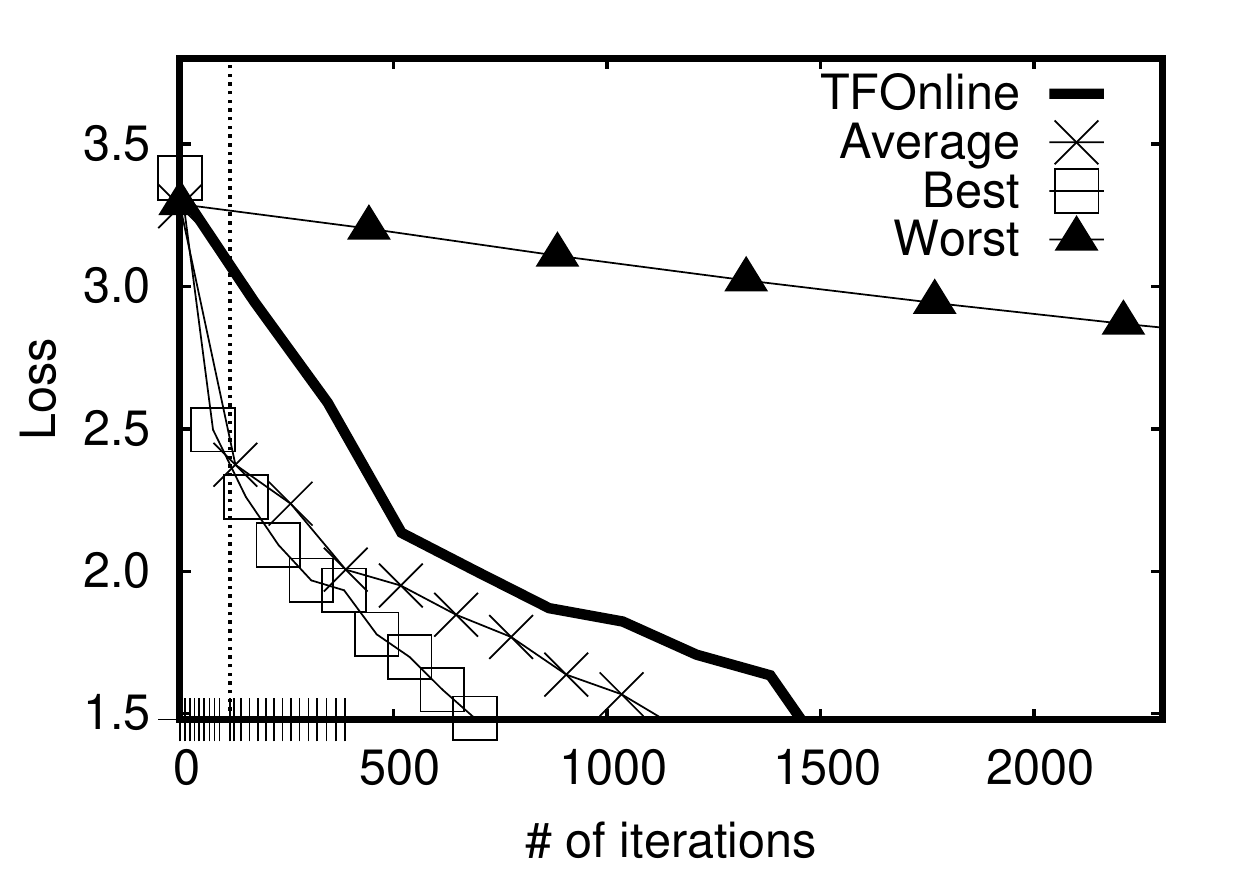}
    }
\caption{Loss vs. Iterations}\label{fig:loss-iter}
\end{center}
\end{figure*}

\subsection{Statistical Efficiency versus Hardware Efficiency}
The completion time of a ML job is a complex interplay between statistical efficiency and hardware efficiency
because a setting good at one efficiency might be a bad setting overall.
Figure \ref{fig:loss-iter} confirms that.  
The figure shows the loss of the jobs with respect to the iterations executed.
In fact, both {\sf \name} and \best have chosen settings that need slightly more iterations to convergence on LogR and CNN (CIFAR) when compared with settings chosen by  \worst and {\sf Average}.
But Table \ref{hvss}, which gives the details of the number of iterations and the execution time per iteration on all workloads,
shows that the settings chosen by {\sf \name} and \best essentially have much better hardware efficiencies
than the settings chosen by \worst and {\sf Average}.
That explains why {\sf \name} and \best have much better end-to-end completion time overall.
On ImageNet8, {\sf \name} has chosen a setting which is more hardware efficient but \best has chosen a setting which is more statistical efficient.  
We believe that is caused by the fact that our estimation function is only a heuristic when facing non-convex problems.
But we remark that the setting chosen by {\sf \name} is a fairly good one after all.

\begin{table*}
\caption{Hardware Efficiency vs. Statistical Efficiency}\label{hvss}
    \centering
    \small
    \scalebox{0.9}{
    
\begin{tabular}{c|cc|cc|cc|cc|}
\cline{2-9}& \multicolumn{2}{c|}{\sf Worst} & \multicolumn{2}{|c|}{\sf Average} & \multicolumn{2}{|c|}{\name} & \multicolumn{2}{|c|}{\sf Best}   \\
\cline{2-9}& \# of & time per  &  \# of  & time per &  \# of  & time per &  \# of  & time per \\
&  iterations  & iteration  &  iterations  &  iteration &  iterations  & iteration  &  iterations  &  iteration \\\hline
\multicolumn{1}{|l|}{LogR}  &      14899  &  0.846s    & 14795    &  0.217s  &      22592  &  0.093s    & 21834    &  0.060s   \\
\multicolumn{1}{|l|}{SVM}   &      106323  &  0.691s    & 227125    &  0.034s  &      223519  &  0.013s    & 225085    &  0.010s    \\
\multicolumn{1}{|l|}{CNN on CIFAR}  &      35426  &  0.157s    & 37520    &  0.023s  &      44827  &  0.011s    & 43601    &  0.005s     \\
\multicolumn{1}{|l|}{CNN on ImageNet}  &      3975  &  24.921s    & 1163    &  6.132s  &      1555  &  3.463s    & 691    &  3.747s     \\\hline
\end{tabular}

    }
    
\end{table*}



Tables \ref{svmsetting}, \ref{logRsetting}, \ref{cifarsetting}, and \ref{imagenetsetting}
list the system settings 
chosen by {\sf Worst}, {\sf Average}, {\sf \name} and {\sf Best} in detail.
For {\sf Average}, the reported setting is the one whose completion time closest to the average completion time.
For {\sf \name}, the reported setting is the final one found by {\sf \name} in the online tuning phase.
Take the settings reported in the SVM experiment (Table \ref{svmsetting}) as an example,
we see that  {\sf \name} found a setting quite close to the {\sf Best}, especially on the server-worker ratio and on the use of parallelism,
which justifies {\sf \name} near-optimal performance in SVM.
When we look at the settings reported in the CNN ImageNet8 experiment (Table \ref{imagenetsetting}), 
we observe {\sf \name} and \best really chose quite different system settings, in which {\sf \name} has chosen a more hardware efficient one but \best has chosen a more statistical efficient one.  
Nonetheless, the setting chosen by  {\sf \name}  is good enough, and outperforms the one chosen by {\sf Average} in terms of end-to-end completion time. 

\begin{table}
	\caption{System Setting in SVM Experiments}\label{svmsetting}
    \centering
    \scalebox{0.8}{
    \footnotesize
    \begin{tabular}{@{}|@{}p{4.5cm}||c|c|c|c|} 
    \hline  
    Knob (names simplified) & {\sf Worst} & {\sf Average} & {\sf TFOnline} & {\sf Best} \\\hline\hline
        \emph{ps} & 35 & 25 & 5 & 2 \\
        \emph{worker} & 1 & 11 & 31 & 34  \\
        \emph{intra\_op\_parallelism\_threads} & 3 & 9 & 1 & 6  \\
        \emph{inter\_op\_parallelism\_threads} & 13 & 7 & 15 & 10  \\
        \emph{do\_common\_subexpression\_elimination} & False & True & False & True  \\
        \emph{max\_folded\_constant\_in\_bytes} & 35500001 & 62500000 & 10485760 & 70500000  \\
        \emph{do\_function\_inlining} & False & False & True & True  \\
        \emph{global\_jit\_level} & ON\_1 & ON\_1 & ON\_1 & OFF  \\
        \emph{infer\_shapes} & False & False & False & True  \\
        \emph{place\_pruned\_graph} & True & False & False & False  \\
        \emph{enable\_bfloat16\_sendrecv} & True & True & False & False  \\
    \hline
\end{tabular}
         
    }

\end{table}

\subsection{Reconfiguration}
In order to evaluate our proposed reconfiguration scheme, especially ODMR,
we carried out a set of experiments on {\sf \name} whose reconfiguration 
is implemented using the baseline method (i.e., checkpointing and recovery).
Table \ref{reconfiguration_table} shows the details about the reconfiguration costs between the two implementations.
Column (a) shows that our reconfiguration scheme reduces the reconfiguration overheads by 400\% (LogR) to 640\% (CNN on CIFAR).
Column (b) shows the average cost of a single reconfiguration.
It shows that our reconfiguration scheme reduces each reconfiguration overhead by 380\%  (LogR) to 760\% (CNN on CIFAR).
The reason of the reconfiguration cost being higher in LogR than the others because the model size of LogR is much bigger than the others (see Table \ref{datasets}).
That influences both the baseline's state-checkpointing-and-recovery cost and ODMR's relocation cost.
Nonetheless, we observe that the number of reconfigurations that took place is actually a fairly small number, between 24 and 50.
Consequently, those costs are worth and offset by the use of better system settings in subsequent iterations,
which our overall experimental results confirm that.


\begin{table*}
\caption{Reconfiguration Cost} \label{reconfiguration_table}
\footnotesize
\centering

\begin{tabular}{|c|c|c|c||c|c|}\hline
Workload & \# of Reconfig & \multicolumn{2}{c||}{(a) Total Overhead}  & \multicolumn{2}{c|}{(b) Overhead per reconfiguration} \\\cline{3-6}
 &  & {\sf Baseline}  & {\sf \name}  &  {\sf Baseline}  & {\sf \name} \\\hline\hline
LogR  &      37  &  1739s    & 444s    &  47s  &      12s        \\
SVM   &      50  &  650s    & 100s    &  13s  &      2s      \\
CNN on CIFAR  &      26  &  416s    & 52s    &  16s  &      2s       \\
CNN on ImageNet8  &      24  &  960s    & 144s    &  40s  &      6s     \\\hline
\end{tabular}

\end{table*}

\subsection{Estimation Quality}
Lastly, we try to understand the quality of our estimation function.
Since our primary goal is not the estimation accuracy as discussed, 
we evaluate the \emph{rank}  \cite{Xu2016} of our estimation function instead.
Specifically, there is a perfect ranking of 100 system settings obtained in the baseline experiments,
where the one with the best completion time, i.e., {\sf Best}, is rank 1st, and {\sf Worst}, has rank 100-th.
This perfect ranking can serve as an oracle to the evaluation.

Consider a random system setting $X_i$ in our baseline experiments.
We segment its execution metrics for every $a$ consecutive pairs of $\langle j, \yb \rangle$.
Then, for each segment we follow our methodology in Section \ref{progress} to form a series of estimation functions $\{r_1, r_2, \dots\}$.
Next, we feed the same convergence threshold to $\{r_1, r_2, \dots\}$ to obtain a series of estimated remaining completion times $\{Y_1, Y_2, ...\}$ with respect to iteration $a$-th, $2a$-th, so on and so forth.
For 100 random settings used in our baseline experiments, we then obtain a table of estimation results like this (note: the numbers below are for illustrations only):

\begin{center}
\begin{small}
\begin{tabular}{c|c|c|c}\hline
Setting & Iteration $60$ & Iteration $120$ & ...  \\
                           & Est. Remaining Time   &  Est. Remaining Time & ...  \\\hline\hline
$X_1$                & 5555  &  3333 & ...  \\
$X_2$                &  4444  &  4222 & ...             \\
...  & ...  & ...   & ...  \\
$X_{100}$                & 7777  &  6666 & ...   \\\hline
\end{tabular}
\end{small}
\end{center}

With a table of estimation results, we can evaluate the quality of the estimation function directly.
Specifically, we can deduce which setting is the estimated ``optimal'' according to the estimation function alone.
For example, as of the moment of iteration 60, our estimation function would regard $X_2$ as the estimated optimal if its estimated remaining time is the lowest among the others.
Similarly, as of the moment of iteration 120, our estimation function would regard $X_1$ as the estimated optimal if its estimated remaining time is the lowest.
Now, consider the moment of iteration 60 and assuming $X_2$ is the estimated optimal setting at that moment, 
we can quantify whether the estimation is reliable by cross-checking the rank of $X_2$ with respect to the oracle.
Concretely, if $X_2$ is also rank 1st in the oracle, that means the estimation is perfect enough to suggest the real optimal.
In contrast, if $X_2$ turns out to rank 100-th in the oracle, that means the  estimated ``optimal'' setting turns out is the worst one among the 100 settings.
The notion of rank based on an actual oracle has been used in \cite{Xu2016} and it was shown that it is way more informative than using the notion of error when evaluating the quality of an estimation function.
So now, for each segment (iterations 1 to 60 is segment 1, iterations 61 to 120 is segment 2, etc.),
we can identify the rank of the estimated optimal in that segment.
We report the average rank of those estimated optimals cross all segments.
Semantically, that is the quality of our estimates \emph{across every possible reconfiguration point} of the training job under {\sf \name}.

Table \ref{ranks} shows the average ranks of our estimation on the four different workloads.
Our estimation function has excellent quality  in LogR and SVM, 
in which its estimated optimals are the third (3.3) and the second (2.0) best settings in real.
As a heuristic for non-convex CNN, 
our estimation function, though not as promising as on LogR and SVM, is still able to return good but not excellent settings that rank within the top-22 percentile.
As an initial prototype,
we regard that as good enough as we can see from the previous experiments (Figure \ref{fig:completion-time}) that  {\sf \name} can
successfully avoid disastrous settings that would have been about  6 times (LogR), 25 times (SVM), 10 times (CIFAR) and 18 times (ImageNet8) slower.
Nonetheless, we are aware of new convergence bounds for two layers neural networks have just been released \cite{ConvergenceAnalysis}.
We will try those new bounds as heuristics estimation function in {\sf \name} for CNN problems in the future.


\begin{table}[t]\small
\caption{Estimation Function Reliability} \label{ranks}
\scalebox{0.9}{

\begin{tabular}{|c||c|c|c|c|}\hline
Workload & LogR & SVM & CNN on CIFAR & CNN on ImageNet8\\ \hline\hline
Rank & 3.3 & 2.0 & 22.0 & 13.0 \\ \hline
\end{tabular}
    
}
\end{table}

\section{Related Work}\label{related}

The (short) history of PS architecture began with systems that were specifically designed for LDA topic modeling \cite{SmolaN10pvldb} and deep network \cite{adam, DeanCMCDLMRSTYN12nips}.
Afterwards, general-purpose ML systems also adopt the PS architecture \cite{LiAPSAJLSS14osdi,XingHDKWLZXKY15kdd}.
Compared with auto-tuning database systems, auto-tuning ML systems is in infancy.
In \cite{KDD2015}, an offline tuner specifically designed for Adam \cite{adam}, a close-source ML system, was presented.
The work manually established an analytical cost-model based on Adam's architecture and design.
Similarly, in \cite{sigmod15resourceelastic}, an offline resource tuner for ML systems was discussed. 
That work, however, focused only on hardware efficiency.
%
Latest works \cite{gdoptimizer} and \cite{nguyen2017reinforcement} discusses the automatic selection of different GD algorithms by manually creating an analytical cost-model
and the automatic placement of operators on CPU/GPU using reinforcement learning, respectively. 
Our scope is way broader than only those.
More importantly, we target online tuning, i.e., a job is executed using better and better system settings as it proceeds.
In contrast, \cite{gdoptimizer} targets offline tuning --- first decide on which GD algorithm to use  and never change that even though a job may last for hours or weeks.
In \cite{Hemingway}, experiments  show that changing the cluster resources online could influence the completion time of ML jobs, which supports the arguments of this paper.
FlexPS \cite{flexps} shares the same vision as us. 
However, FlexPS only supports only one knob -- the worker-server ratio. Furthermore, it requires users to learn a completely new programming model and API.  In contrast, our techniques in this paper are general.  We have shown that we can apply our techniques to the popular {\sf Tensorflow} and users can enjoy better efficiency with no pain.

There are distributed systems specialized for deep learning, for example, SINGA \cite{singa},
MXNET \cite{MxNet}, FireCaffe \cite{FireCaffe}, SparkNet \cite{SparkNet}, Omnivore \cite{Om2016}, and Project Adam \cite{adam}.
Decoupling hardware and statistical efficiency is not new there. 
For example, MXNet reported hardware efficiency and statistical efficiency separately,
SINGA studied their tradeoff,
and Omnivore leveraged that tradeoff to improve end-to-end running time.
However, they have not studied the issues of \emph{online tuning} and \emph{system reconfigurations} as we do.

In machine learning, auto hyper-parameter tuning 
(e.g., tuning the number of the hidden layers in a deep neural network)
that finds the best model is an automated machine learning (AutoML) problem \cite{automl}.  
State-of-the-art hyper-parameter tuning 
systems \cite{hyperband, BOHB, massiveparellel, SparksTFJK15, KotthoffTHHL17, GolovinSMKKS17, Becker-KornstaedtSZ00, gpyopt2016, autosklearn, easeml} 
however have not addressed online tuning that includes system parameters.


Performance modeling and progress estimation are interesting problems in their own right.
For example, Ernest \cite{VenkataramanYFR16} trains a performance model for machine learning applications.
However, Ernest and more recent work  \cite{JustusBBM18} have only put the estimation of statistical efficiency as a future work.
Progress indicator is a useful add-on in analytical systems because  that lets users know when will they obtain the results \cite{PELI, CIDRLI, EPSIGMOD2004, TPISIGMOD2004, MapReduceICDE2010}.
In this paper, we have pioneered the first progress indicator for ML systems 
through giving initial solutions to the statistical progress estimation problem.

\section{Conclusion and Future Work}
\label{conclusion}

In this paper, we make a case for building a parameter-server system prototype that supports self-tuning.
We show that the performances of
machine learning (ML) systems, like database systems,
are also subjected to the values of system parameter.
However, unlike database systems, ML systems can afford online on-job training and tuning because of the long-running nature of ML iterative programs.
To this end, we have developed an online optimization framework that is suitable to all ML systems.
We have also developed initial solutions to approach the online statistical progress estimation problem.
Furthermore, we have developed new techniques to carry out online reconfiguration in a lightweight and non-quiescent mode.
As an initial effort to showcase our techniques,
we have implemented a prototype on top of {\sf TensorFlow}.
Experiments show that various ML tasks gain speedup by a factor of 1.4$\times$ to 18$\times$.

As a prototype, {\sf \name} has included only one specific statistic progress estimation function. 
Although empirical results show that it works well even on workloads that violate its assumption,
we are going to devise specific estimation functions for each kind of ML problems and algorithms.
From the system perspective, we plan to extend our idea to other system architecture (e.g,. the peer to peer architecture on a scale-up machine \cite{peer2peer}),
and to platforms with heterogenous machines (e.g., \cite{Ce2017}).
We are also in the process of using \emph{transfer learning} \cite{tl} to eliminate the initialization phase.
Specifically, when the framework receives a new ML job $J$,
it shall search the repository and locate a previous job $\hat{J}$ that is most similar to $J$.
Then it shall transfer all candidate settings $\mathcal{X}_{\hat{J}}$ that $\hat{J}$ had ever picked
to be $J$'s candidate settings.
In this case, determining the value for $b$, the number of initial settings to try during the initialization phase, is no longer a question.
OtterTune  \cite{andyMLtune} has also leveraged a similar idea  when facing new DB workloads.
We believe the formal use of  {transfer learning} on ML system tuning would be promising.

\bibliographystyle{IEEEtran}
\bibliography{1-ml,1-tuning,1-theory}

\appendices
\section{Notations}
\label{Appendix}
We summarized all used notations in Table~\ref{tbl:notations}.
\begin{table}[h] \small
\caption{Notation table}
\begin{tabular}{|c|p{0.73\linewidth}|}\hline
\bf Notation & \bf Meaning \\ \hline
$X$ &  system setting \\ \hline
$X^*$ & optimal or near-optimal system setting \\ \hline
$\loss_i$ & the loss of the first iteration of using setting $X_i$\\ \hline
$\langle X, \loss \rangle$ &  $(d+1)$-dimensional vector that includes both the system setting values and the loss of the model \\ \hline
$\T(\langle X, \loss  \rangle)$ & the remaining completion time of the job  if we switch to setting $X$ where the model has reached a loss down to $\loss$ \\ \hline
$\ya$ & the execution time of $j$-th iteration \\ \hline
$\yb$ & the loss of iteration $j$ under setting $X_i$\\ \hline
$Y_i$ & the estimated remaining completion time at setting $X_i$\\ \hline
$\langle j, X_i, \ya, \yb \rangle$ & collected execution metrics \\ \hline
$\langle X_i, \loss_i, Y_i \rangle$ & training data for the Bayesian Optimization (BO) \\ \hline
$\w^j$ & model parameters after $j$-th iteration \\ \hline
$\w{}^*$ & the optimal model parameters \\ \hline

\end{tabular}
\footnotesize
\label{tbl:notations}
\end{table}


\section{Implementing {\sf \name}}\label{app:B}

We have implemented our techniques on top of {\sf TensorFlow} and name our prototype {\sf \name}.

\subsection{User-Program}

Currently {\sf TensorFlow} exposes all system settings through the class constructors and class attributes of the core classes.
Table \ref{tab:\name} (left) shows how ML users specify those settings within the program.
We have implemented a Python module for {\sf \name} so that users no longer need to specify the system settings anymore.
Table \ref{tab:\name} (right) shows the corresponding program with {\sf \name} installed.
We can see that a ML user no longer needs to bother for those system settings,
except she is required to implement her {\sf TensorFlow} program by extending a new class {\sf MLJobFramework} provided by \name.
That class is implemented in the front-end to collect runtime statistics.

\begin{table*}[h!]
	\caption{Example of {\sf \name} module}\label{tab:\name}
	\centering
	\scalebox{0.9}{
        \includegraphics[width=15cm]{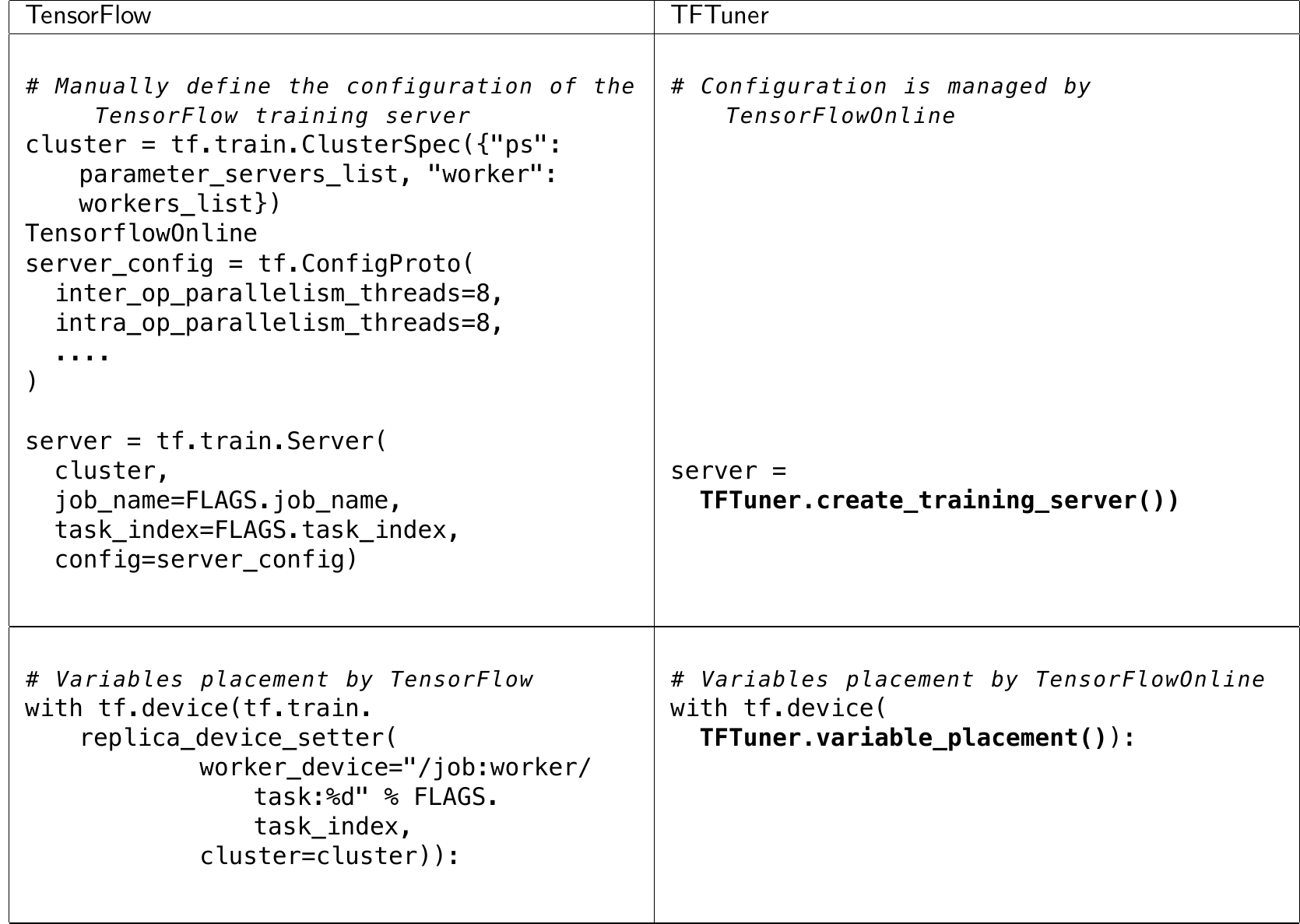}
    }
\end{table*}

\subsection{Front-end}

{\sf TensorFlow} existing implementation already has explicit facilities to implement CKP, MDR, and TDR.
Specifically, CKP can be invoked by {\sf Saver.save()}, 
MDR can be invoked by {\sf Saver.restore()}, 
and TDR can be invoked by reading training data through  {\sf TensorFlow} {\sf tf.ReaderBase} with HDFS filesystem plugin.

Now, we discuss how we implement ODMR (On-Demand-Model-Relocation) in {\sf TensorFlow}.
The placement of parameters is controlled by the execution graph generated by TensorFlow front-end.
When data reallocation (e.g., changing the number of parameter server) occurs, attribute {\sf tf.Variable::device}, which controls the location of the parameters, is updated according to the parameter mapping generated by \name. 
To push also the original data value under ODMR,
we added an extra operation to the front-end to do so.

\subsection{Back-End}
We modify the back-end of {\sf TensorFlow} in order to reduce the overhead of SSR. Currently, if we want to carry out SSR (e.g., changing of the number of intra\_op\_parallelism\_threads) in {\sf TensorFlow}, the whole {\sf TensorFlow} program has to completely restart since the back-end of {\sf TensorFlow}  cannot change the system knobs on-fly.
In {\sf \name}, we modify TensorFlow back-end and expose a new function called {\sf Reconfig()} in the API. 
It allows the back-end to accept new knob values from the front-end without restarting the whole job.


\clearpage
\onecolumn
\section{Additional Experiment Results}

\begin{table}[h!]
    \caption{System Setting in LogR experiment}\label{logRsetting}
    \centering
    \small
    \scalebox{0.8}{
    
    \begin{tabular}{@{}|@{}p{5.8cm}||c|c|c|c|} 
    \hline  
    Knob (names simplified) & {\sf Worst} & {\sf Average} & {\sf TFOnline} & {\sf Best} \\\hline\hline
        \emph{ps} & 34 & 28 & 20 & 18 \\
        \emph{worker} & 2 & 8 & 16 & 18  \\
        \emph{intra\_op\_parallelism\_threads} & 12 & 10 & 6 & 11  \\
        \emph{inter\_op\_parallelism\_threads} & 11 & 6 & 10 & 5  \\
        \emph{do\_common\_subexpression\_elimination} & False & True & True & True  \\
        \emph{max\_folded\_constant\_in\_bytes} & 86954782 & 10485760 & 25092366 & 28952477  \\
        \emph{do\_function\_inlining} & False & True & False & True  \\
        \emph{global\_jit\_level} & ON\_2 & OFF & ON\_1 & ON\_2  \\
        \emph{infer\_shapes} & True & True & True & True  \\
        \emph{place\_pruned\_graph} & True & False & True & False  \\
        \emph{enable\_bfloat16\_sendrecv} & False & False & True & True  \\
    \hline
\end{tabular}
        
    }
\end{table}

\begin{table}[h!]
\caption{System Setting in CNN on CIFAR Experiments}\label{cifarsetting}

    \centering
    \small
        \scalebox{0.8}{

    \begin{tabular}{@{}|@{}p{5.8cm}||c|c|c|c|} 
    \hline  
    Knob (names simplified) & {\sf Worst} & {\sf Average} & {\sf TFOnline} & {\sf Best} \\\hline\hline
        \emph{ps} & 33 & 16 & 6 & 5 \\
        \emph{worker} & 3 & 20 & 30 & 31  \\
        \emph{intra\_op\_parallelism\_threads} & 2 & 2 & 3 & 12  \\
        \emph{inter\_op\_parallelism\_threads} & 14 & 14 & 13 & 4  \\
        \emph{do\_common\_subexpression\_elimination} & False & True & False & False  \\
        \emph{max\_folded\_constant\_in\_bytes} & 5125478 & 10485760 & 65136941 & 50785965  \\
        \emph{do\_function\_inlining} & False & True & False & True  \\
        \emph{global\_jit\_level} & ON\_2 & OFF & ON\_1 & OFF  \\
        \emph{infer\_shapes} & True & True & True & True  \\
        \emph{place\_pruned\_graph} & False & False & True & True  \\
        \emph{enable\_bfloat16\_sendrecv} & True & False & True & False  \\
    \hline
\end{tabular}
        
}
\end{table}

\begin{table}[h!]
\caption{System Setting in CNN on ImageNet8 Experiment}\label{imagenetsetting}

    \centering
    \small
        \scalebox{0.8}{

    \begin{tabular}{@{}|@{}p{5.8cm}||c|c|c|c|} 
    \hline  
    Knob (names simplified) & {\sf Worst} & {\sf Average} & {\sf TFOnline} & {\sf Best} \\\hline\hline
        \emph{ps} & 31 & 25 & 27 & 13 \\
        \emph{worker} & 5 & 11 & 9 & 23  \\
        \emph{intra\_op\_parallelism\_threads} & 12 & 1 & 1 & 12  \\
        \emph{inter\_op\_parallelism\_threads} & 4 & 15 & 15 & 4  \\
        \emph{do\_common\_subexpression\_elimination} & False & True & False & True  \\
        \emph{max\_folded\_constant\_in\_bytes} & 96500000 & 10485760 & 10485760 & 10485760  \\
        \emph{do\_function\_inlining} & True & True & False & True  \\
        \emph{global\_jit\_level} & ON\_2 & OFF & ON\_2 & OFF  \\
        \emph{infer\_shapes} & False & True & False & True  \\
        \emph{place\_pruned\_graph} & False & False & False & False  \\
        \emph{enable\_bfloat16\_sendrecv} & False & False & False & False  \\
    \hline
\end{tabular}
            
}
\end{table}

\end{document}